\documentclass[11pt,a4paper]{article}
\usepackage[a4paper, margin=0.6in]{geometry}
\usepackage[T1]{fontenc}
\usepackage{lmodern}
\usepackage{listings}
\usepackage{subcaption}
\usepackage[section,above]{placeins}
\usepackage[numbers,sort&compress]{natbib}
\usepackage[table]{xcolor}
\usepackage{float}
\newfloat{algorithm}{htbp}{loa}
\floatname{algorithm}{Algorithm}
\usepackage{booktabs, array}
\usepackage{tabularx}
\usepackage{longtable}
\usepackage{multirow}
\usepackage[hyperfootnotes=false]{hyperref}
\hypersetup{hidelinks}
\usepackage{makecell}
\usepackage{graphicx}
\usepackage{amsmath,amssymb,amsfonts}
\usepackage{amsthm}
\usepackage{mathrsfs}
\usepackage[title]{appendix}
\usepackage{textcomp}
\usepackage{manyfoot}
\usepackage{threeparttable}
\usepackage{lineno}
\usepackage{authblk}
\usepackage{enumitem}
\usepackage{caption}
\usepackage{tikz}


\author[1,*]{Mellow Baixuan Chen}
\affil[1]{Courant Institute of Mathematical Sciences, New York University, New York, 10012, NY, USA}
\author[1,2]{Alan A. Kaptanoglu}
\affil[2]{William E. Boeing Department of Aeronautics and Astronautics, University of Washington, Seattle, 98195, Washington, USA}
\date{}
\title{Machine-learned spectral PDE solvers for fluids and plasmas}

\begin{document}
\maketitle
\begingroup
\renewcommand\thefootnote{}\footnotetext{$^{*}$\,Corresponding author: bc3281@nyu.edu}
\endgroup
\vspace{-3.5em}
\begin{abstract}
Sparse regression methods for discovering governing partial differential equations (PDEs) from data are sensitive to observation noise and are not constrained to reproduce the data under forward integration. We address both limitations by performing weak-form discovery in the spatial discretization in which the PDE is solved, learned by differentiating through multistep rollouts anchored at the observed states. The new method, Weak-form Anchored Rollout for PDEs (WARP), produces a machine-learned PDE solver. In this work, we focus on systems amenable to a spectral Galerkin solver. Across eight benchmark fluid and plasma systems, including high-dimensional turbulent simulations, WARP produces numerically stable rollouts from new random initial conditions and substantially improves coefficient recovery, rollout accuracy, and statistical measures compared with a single-step weak-form SINDy baseline.
\end{abstract}

\section{Introduction}
Discovering governing partial differential equations (PDEs) from spatiotemporal data is a central task in scientific machine learning. It is also a difficult inverse problem. A finite trajectory visits only a small part of state space, candidate operators can be strongly correlated along that path, and distinct equations can fit the same local data. Measurement noise, differentiation error, missing fields, discretization bias, and unresolved closure terms further confound the coefficients. For nonlinear and chaotic systems, small structural or coefficient errors feed back through the dynamics and eventually prevent accurate forecasts over sufficiently long time horizons. Successful identification therefore requires information-rich training data, an identifiable representation, a suitable closure, and a learned model that remains usable under forward integration~\citep{Brunton2016,Rudy2017,KaptanogluBenchmark2023,VASEY2025113719}.

Traditional SINDy or other single-step prediction (derivative fitting) methods present a second problem beyond noise sensitivity. A small instantaneous residual does not enforce dissipation, invariants, boundedness, or correct behavior away from the observed states. Small spurious terms can create anti-diffusion or energy production, so models that fit derivatives accurately may diverge when integrated from the training state or a new initial condition~\citep{LoiseauBrunton2018,KaptanogluMHD2021,KaptanogluTrapping2021,GoyalBenner2022,LeeTraskStinis2022,BannaUnrolled2027}; Appendix~\ref{app:longroll} tests this on our benchmarks. Other approaches enforce stability through physics constraints, Lyapunov conditions, bounded quadratic model classes, or stability-promoting regularization~\citep{SchlegelNoack2015,KaptanogluTrapping2021,LeeTraskStinis2022,GoyalDuffBenner2023,RodriguezLyaNet2022}.

We combine weak spatial projection with solver-in-the-loop coefficient training. The unknown PDE is discretized as a spectral Galerkin solver, using an orthonormal Fourier basis to represent the state and test the weak form. The resulting modal ordinary differential equation (ODE) is integrated from observed anchor states, and its candidate coefficients are optimized by differentiating through adaptive Dormand--Prince rollouts. We call the method Weak-form Anchored Rollout for PDEs (WARP). Its output consists of explicit PDE coefficients and an autonomous solver in the same finite-dimensional weak formulation used during training. We emphasize that the foundation presented here (i.e. identifying the equations by direct forward integration in a Galerkin formulation) can be extended to arbitrary PDEs and boundary conditions, different test function bases, and a variety of ODE integrators.

We will motivate WARP by applying it to increasingly complex periodic fluid and plasma systems, from scalar nonlinear and dispersive equations to coupled turbulent and magnetohydrodynamic (MHD) dynamics. The physical role and numerical setup of each benchmark are introduced with its results in Sec.~\ref{sec:experiments}. Every fit uses a single observed training trajectory. This is a different data regime from large multi-trajectory neural-operator benchmarks~\citep{li2021fourier,PDEBench2022}, although rollout-based operator inference also explicitly addresses scarce and noisy data~\citep{UyHartmannPeherstorfer2023}. Across eight systems, WARP improves coefficient recovery and, typically, autonomous rollout accuracy over a single-step weak-formulation comparison. Because both methods use the same weak spatial projection, the comparison isolates the benefit of multistep rollout.

\section{Background and Motivation}
\label{sec:background}
Data-driven PDE methods differ both in what they learn and in how they train it. Sparse library regression and symbolic methods recover equations with named operators and numerical coefficients~\citep{Brunton2016,Rudy2017,Schaeffer2017,MESSENGER2021110525}. Neural methods can instead learn a vector field, a numerical stencil, a correction to an existing solver, or a map between functions~\citep{NeuralODE,LongPDENet2018,li2021fourier,DeepONet}. These categories overlap: PDE-Net 2.0, for example, uses a neural architecture to recover a symbolic PDE~\citep{LongPDENet2019}. WARP estimates the coefficients of specified differential operators, without a neural network in the right-hand side. The more specific comparison is therefore with methods that identify their model by repeatedly integrating that same model during training.

Neural ODEs establish this training principle for a network-parameterized vector field: the loss compares observations with a numerical solution, and gradients pass through the integration~\citep{NeuralODE}. Latent and augmented neural ODEs, universal differential equations, stiff-system formulations, and Lyapunov-based training extend this approach~\citep{rubanova2019latentodesirregularlysampledtime,dupont2019augmentedneuralodes,Kidger2022,Rackauckas2020,KimStiff2021,RodriguezLyaNet2022}. A learned right-hand side is evaluated on its own evolving state at the solver stages. This is also the consistency condition used by WARP; it is distinct from evaluating a candidate library only on measured states and subsequently integrating the fitted model.

A particularly close precedent is operator inference with rollouts~\citep{UyHartmannPeherstorfer2023}. Classical operator inference estimates reduced operators with the polynomial structure of a projected governing equation~\citep{PeherstorferWillcox2016}. Its rollout extension optimizes those operators through multistep trajectories, restarts the trajectories at projected observations, and uses the trained reduced solver for prediction. Thus, learning an interpretable projected dynamical system through observation-anchored integration already has a direct precedent. WARP restricts the unknowns to coefficients of named physical-space differential operators, whose weak projections assemble the modal right-hand side. These shared PDE coefficients tie together the equations for all retained modes, rather than estimating general reduced polynomial operator arrays. This distinction concerns the parameterization and spatial weak formulation, not the use of rollouts itself.

Sparse-identification methods also incorporate integration into training. Automatic-differentiation SINDy jointly estimates dynamics and observation noise using short predictions~\citep{Kaheman2022}; SINDy with structure constraints uses trajectory information~\citep{LeeTraskStinis2022}. Runge--Kutta SINDy fits an integration map~\citep{GoyalBenner2022}, and Unrolled-SINDy evaluates the candidate library on intermediate predicted states between observations, using multiple internal Euler or Runge--Kutta steps~\citep{BannaUnrolled2027}. PDE-Net learns spatial differentiation filters and an evolution model through repeated time-stepping blocks~\citep{LongPDENet2018}; PDE-Net 2.0 adds a symbolic response network~\citep{LongPDENet2019}. These are relevant examples of learning a deployable solver from its own predictions. WARP instead fixes analytic basis derivatives and constructs the learned PDE through weak spatial projection, then optimizes its coefficients through adaptive multistep integration.

Learned discretizations and closures provide another close comparison. Data-driven discretization learns spatial approximations from resolved simulation data~\citep{BarSinai2019}. Kochkov et al. train corrections within a differentiable fluid solver using losses over repeated steps~\citep{Kochkov2021}, and Solver-in-the-Loop differentiates through a solver coupled to learned corrections~\citep{UmSolverLoop2020}. The deep learning PDE augmentation method trains neural closures through the governing PDE and its adjoint, rather than fitting closure targets alone~\citep{SirignanoDPM2020}. Kang and Constantinescu learn subgrid source terms through integration of a coarse method-of-lines model~\citep{KangConstantinescu2023}. Most directly relevant to the weak formulation, Jung and Constantinescu learn state-dependent corrective coefficient fields inside a finite-element variational solve and optimize a multistep rollout loss~\citep{JungConstantinescu2026}. These methods repeatedly evaluate the learned components on predicted states and deploy the resulting corrected solver. Their target is an effective correction to a specified coarse model; WARP's present target is a small vector of explicit PDE coefficients. The distinction becomes less sharp for the projected-DNS experiment, where fitted coefficients can also absorb unresolved-scale effects.

Weak formulations transfer derivatives from measurements to test functions and have been developed for ODE and PDE identification, online estimation, convergence analysis, ensemble selection, laboratory flows, and plasma applications~\citep{Messenger2021,MESSENGER2021110525,Gurevich2019,Reinbold2020,pmlr-v190-a-messenger22a,Russo2024,EnsembleSINDy,Fasel2025,Fasel2026lowgoactivelearning,Reinbold2021,Gurevich2024SPIDER,AlvesFiuza2022,MessengerStrong2024}. Moving-horizon optimization offers a related way to combine identification with integration over observation windows~\citep{Lejarza2022}. Weak projection alone does not remove all statistical bias: both the response and the library depend on noisy observations. WENDy treats this errors-in-variables structure through covariance-aware iterative reweighting~\citep{BortzWENDy2023}, while the consistency analysis of WSINDy identifies noise-dependent biases from nonlinear library evaluation~\citep{MessengerBortzConsistency2025}. WARP avoids a finite-difference time derivative, but its noisy anchors and targets do not make it an unbiased estimator.

WARP therefore combines established rollout training with a specific interpretable spatial discretization: the same orthonormal Fourier functions represent the state and test the PDE, and their projected candidate operators define the solver used during optimization and deployment. On periodic domains this removes boundary terms and avoids separate localized-test-function placement and shape choices~\citep{CanutoSpectral2006}. We do not claim that learning a solver through self-consistent integration, or combining weak forms with rollout training, is new by itself. Our contribution is this construction for explicit PDE-coefficient identification and its controlled comparison with derivative fitting on the identical modal library.

Plasma applications include sparse MHD models, kinetic-to-fluid discovery, plasma-boundary control models, and reduced models of oscillations and turbulence~\citep{KaptanogluMHD2021,AlvesFiuza2022,LoreSOLPS2023,DonaghuGermaschewski2023,VASEY2025113719,Thakur2022,BayonBujan2024,KaptanogluPoP2023,GahrFarcasJenko2024}. Related discovery of continuum equations and closures from particle data also extends to multiphase flows, mean-field systems, and active matter~\citep{Beetham2021,MessengerMeanField2022,Supekar2023}. The low-dimensional-chaos benchmark of \citet{KaptanogluBenchmark2023} provides a broad comparison of sparse identification algorithms, including the benefits of weak formulations.

\section{Methodology}\label{sec:Methodology}
\subsection{Spectral Galerkin discretization}\label{sec:galerkin}
Let $u(x,t)\in\mathbb R^{N_f}$ be an $N_f$-field state on the periodic domain $\Omega=[0,L]^s$, with spatial dimension $s=1$ or $2$. The candidate PDE is
\begin{equation}
\partial_t u=\mathcal D(u)\xi
=\sum_{\ell=1}^{P}\xi_\ell\mathcal D_\ell(u),
\qquad \xi\in\mathbb R^P,\quad
\mathcal D(u)(x,t)\in\mathbb R^{N_f\times P}.
\label{eq:generic_pde}
\end{equation}
Here $\mathcal D_\ell$ is a specified differential operator, $r$ is the largest spatial derivative order in the library, and $\mathcal D(u)$ has columns $\mathcal D_\ell(u)$. A term that appears in only one field equation has zero entries in the other components. The scalar coefficients in $\xi$ are learned independently unless the library explicitly ties them together.
In one dimension we use the orthonormal real Fourier basis
\begin{align}
\psi_0(x)&=L^{-1/2},\quad
\psi_{2m-1}(x)=\sqrt{2/L}\cos(2\pi m x/L),\quad
\psi_{2m}(x)=\sqrt{2/L}\sin(2\pi m x/L),
\label{eq:fourier_basis}
\end{align}
for $m=1,\ldots,M$, $\psi(x)\in\mathbb R^K$,
where $\psi=(\psi_0,\ldots,\psi_{K-1})$ and each $\psi_k$ is scalar-valued. In two dimensions we use tensor products of these functions. The retained Fourier \emph{band} is the finite set of integer wavevectors with $|m_j|\le M$ in each coordinate $j$; it contains $K=(2M+1)^s$ real basis functions per field. Let $V_M$ be their span and $\mathcal P_M:L^2(\Omega;\mathbb R^{N_f})\to V_M^{N_f}$ the componentwise orthogonal projection. The \emph{projected band} of a field is the part retained by this projection, with all other Fourier coefficients discarded. With the inner product integrated over $\Omega$, the modal coefficients and reconstructed field are
\begin{equation}
q_k(t)=\langle u(\cdot,t),\psi_k\rangle\in\mathbb R^{N_f},
\qquad
u_M(x,t):=\mathcal P_Mu(x,t)=\sum_{k=0}^{K-1}q_k(t)\psi_k(x)\in\mathbb R^{N_f}.
\label{eq:modal_state}
\end{equation}
The inner product in Eq.~\eqref{eq:modal_state} is taken separately for each field. Stacking all $q_k$ gives $q\in\mathbb R^d$, where $d=N_fK$.

We call a data set \emph{band-closed} when it is generated by evolving this finite-dimensional Galerkin system itself; the right-hand side is evaluated from $u_M$ and projected back onto $V_M^{N_f}$ at every stage. No independently evolving discarded modes feed back into the retained state. This does not mean that the finite band is closed under nonlinear multiplication; products can generate higher modes, which are removed by projection and dealiasing. In contrast, projecting a finer DNS trajectory onto the same band generally leaves unresolved interactions in its evolution, so those data are not band-closed.

The basis functions are also the Galerkin test functions~\citep{CanutoSpectral2006}. Uniform trapezoidal quadrature is spectrally accurate for smooth periodic integrands and exponentially accurate when they are analytic~\citep{TrefethenWeideman2014}. Basis derivatives are evaluated analytically.

\subsection{Weak-form modal ODE}\label{sec:modal_ode_sec}
Testing Eq.~\eqref{eq:generic_pde} against the basis and integrating by parts transfers spatial derivatives to known test functions. After concatenating the field and mode indices into $i=1,\ldots,d$, the result is
\begin{equation}
\dot q_i=F_i(q;\xi)=\sum_{\ell=1}^{P}\xi_\ell A_{i\ell}(q),
\label{eq:modal_ode}
\end{equation}
Here $F:\mathbb R^d\times\mathbb R^P\to\mathbb R^d$ and $A:\mathbb R^d\to\mathbb R^{d\times P}$; $A(q)$ is the weak-form library evaluated on the reconstructed field $u_M$. Equation~\eqref{eq:modal_ode} is the semi-discrete model used for data generation, coefficient training, and deployment in the synthetic benchmarks.

\subsection{Differential-operator representations}\label{sec:rhs_representations}
A SINDy-style interpretable model begins with a library of candidate differential operators such as constants, polynomial products of fields, spatial derivatives, and products of fields with derivatives~\citep{Brunton2016,Rudy2017,KaptanogluPySINDy}. Projecting each candidate into the basis produces one column of the modal library. Sparse regression or another selection rule then determines which columns are active and estimates their coefficients. The result is a symbolic PDE whose terms and coefficients can be inspected directly.

A neural representation replaces the fixed library expansion by a parameterized map $F_\theta(q)$, or by a network that constructs spatial operators before projection. This offers greater expressivity when the relevant terms are unknown, at the cost of a coefficient vector that no longer names a compact set of physical operators~\citep{NeuralODE,LongPDENet2019,li2021fourier}. The rollout objective below can optimize either representation. In every experiment reported here, however, we use the exactly correct differential-operator library and estimate all of its coefficients. Fixing the support removes term-selection error and isolates the effect of the training criterion. Larger and misspecified libraries are left to future work, and investigated briefly in Appendix~\ref{app:biglib}.

\subsection{Observation model and noise}\label{sec:noise_model}
Observations of each field on the uniform spatial grid are corrupted independently by additive Gaussian noise,
\begin{equation}
\hat u_a(x_\ell,t_n)=u_a(x_\ell,t_n)+\varepsilon_{a\ell n},\qquad
\varepsilon_{a\ell n}\sim\mathcal N\!\bigl(0,\,(\eta\sigma_a)^2\bigr),\quad a=1,\ldots,N_f,
\label{eq:noise_model}
\end{equation}
where $\sigma_a$ is the standard deviation of field $a$ over the clean trajectory and $\eta$ is the per-field noise-to-signal ratio~\citep{Rudy2017,MESSENGER2021110525}. The modal observations are $q_{k,\mathrm{data}}(t_n)=\langle\hat u(\cdot,t_n),\psi_k\rangle\in\mathbb R^{N_f}$, stacked in the same order as $q$. 

\subsection{Data generation}
\label{sec:consistent_datagen}

Reference trajectories are generated with adaptive Dormand--Prince integration~\citep{DormandPrince1980} and $2/3$ dealiasing of nonlinear products~\citep{Orszag1971}. Optimization rollouts, training-trajectory predictions, and held-out evaluations use the same adaptive integrator and tolerance policy. 
We also perform one experiment with externally generated direct numerical simulation (DNS) data; see Table~\ref{tab:settings}.

\subsection{Single-step derivative-fit comparison}\label{sec:fd_comparison}
The comparison method shares the basis, modal observations, and $A(q)$ with WARP. It evaluates the library on $q_{\mathrm{data}}(t_n)$ and regresses the central finite difference $[q_{\mathrm{data}}(t_{n+1})-q_{\mathrm{data}}(t_{n-1})]/(2\Delta t_{\rm s})$ against it by least squares. We abbreviate this method as FD throughout the remainder of the paper. The comparison therefore isolates multistep rollout from single-step derivative fitting within the same weak formulation.

\subsection{WARP anchored multistep rollout training}\label{sec:warp_training}
WARP re-evaluates $A(q)$ on every predicted state and fits $\xi$ through trajectories of Eq.~\eqref{eq:modal_ode}. Let $H$ be the rollout horizon in snapshot intervals. The nonzero offsets are $\mathcal I:=\{-H,\ldots,-1,1,\ldots,H\}$ for bidirectional training, $\mathcal I:=\{-H_b,\ldots,-1,1,\ldots,H\}$ with a shorter backward horizon $H_b<H$, used with $H_b=5$ for the Hasegawa--Wakatani (HW) system, and $\mathcal I:=\{1,\ldots,H\}$ for forward-only configurations. The selected anchor set is
\[
\mathcal B:=\{j_r\}_{r=1}^{N_{\mathcal B}}\subseteq
\{j\in\{1,\ldots,T\}\mid j+i\in\{1,\ldots,T\}\ \text{for every }i\in\mathcal I\}.
\]
For each $j\in\mathcal B$ and $i\in\mathcal I$, define
$\hat q_{j,i}(\xi):=\operatorname{ODESolve}(\dot q=F(q;\xi),q(t_j)=q_{\mathrm{data}}(t_j),t_{j+i})$. The implemented objective is
\begin{equation}
\mathcal L(\xi):=\frac{1}{Z}
\sum_{j\in\mathcal B}\sum_{i\in\mathcal I}w_i
\left\|\hat q_{j,i}(\xi)-q_{\mathrm{data}}(t_{j+i})\right\|_2^2,
\qquad
Z:=\sum_{j\in\mathcal B}\sum_{i\in\mathcal I}
\left\|q_{\mathrm{data}}(t_{j+i})\right\|_2^2.
\label{eq:bidirectional_loss}
\end{equation}
The denominator $Z$ is fixed after the training trajectory, anchors, and offsets are chosen. For multi-field states with disparate amplitudes the residual norm is evaluated with per-field inverse-root-mean-square (rms) weights (used for the Orszag--Tang vortex, Sec.~\ref{sec:ot2d}). Regularization is often added to Opt.~\eqref{eq:bidirectional_loss} but we omit it because the candidate library of terms is chosen to contain only the correct PDE terms. Training rollouts use JAX \texttt{odeint}~\citep{jax2018github} with adaptive Dormand--Prince integration~\citep{DormandPrince1980}, whose gradients with respect to $\xi$ are obtained by the adjoint sensitivity method. Nonfinite rollouts receive a fixed loss of $10^6$. Limited-memory Broyden--Fletcher--Goldfarb--Shanno (L-BFGS) optimization~\citep{LiuNocedal1989} is used, and the optimizations are initialized from the FD estimate.

\paragraph{Post-training solver.}
Once $\xi$ is learned, deployment requires no future observations. Equation~\eqref{eq:modal_ode} supplies an autonomous spectral Galerkin solver, and adaptive Dormand--Prince integration advances every reported evaluation. Algorithm~\ref{alg:training} summarizes the procedure.

{\setlength{\intextsep}{6pt plus 1pt minus 1pt}
\begin{algorithm}
\centering
\setlength{\fboxsep}{2.5pt}
\fbox{\begin{minipage}{0.96\linewidth}
\fontsize{8pt}{9pt}\selectfont\raggedright
\setlength{\parskip}{0pt}
\setlength{\parindent}{0pt}

\textbf{WARP coefficient training.}
\vspace{0.5pt}\hrule\vspace{1pt}

\textbf{Input.} Observed vector-field snapshots $\{\hat{u}(\cdot,t_n)\}_{n=1}^T$, Fourier basis $\{\psi_k\}_{k=0}^{K-1}$, candidate operators, and rollout offsets $\mathcal I$.

\vspace{1pt}
\textbf{Preprocessing.}
\begin{enumerate}[label=\arabic*., leftmargin=1.5em, itemsep=0pt, topsep=0pt, parsep=0pt]
\item Project each field onto the basis to obtain $q_{\mathrm{data}}(t_n)$.
\item Precompute the basis and its analytic derivatives on the grid.
\item Initialize $\xi$ with the FD weak-formulation estimate.
\end{enumerate}

\vspace{0.5pt}
\textbf{Training loop.}
\begin{enumerate}[label=\arabic*., leftmargin=1.5em, itemsep=0pt, topsep=0pt, parsep=0pt, start=4]
\item Select valid anchors $\mathcal B$ and set each rollout initial state to $q_{\mathrm{data}}(t_j)$.
\item Integrate Eq.~\eqref{eq:modal_ode} to the offsets $i\in\mathcal I$, re-evaluating $A(q)$ at every solver stage.
\item Evaluate Eq.~\eqref{eq:bidirectional_loss}, obtain $\nabla_\xi\mathcal L$ by automatic differentiation of the solve, and update $\xi$ with L-BFGS.
\end{enumerate}
\vspace{0.5pt}
\textbf{Output.} Learned PDE coefficients $\xi$ and an autonomous spectral Galerkin
solver $\dot q=F(q;\xi)$.
\end{minipage}}
\setlength{\abovecaptionskip}{3pt}
\caption{WARP training procedure}
\label{alg:training}
\end{algorithm}}

\subsection{Evaluation metrics}\label{sec:eval_metrics}

We use three seeds $\{42,123,456\}$ in order to vary the observation-noise realization on a fixed training trajectory. We report coefficient error,
\begin{align}
\label{eq:coefficient_error_def}
E_\xi:=\|\xi-\xi^\ast\|_2,
\end{align}
as mean $\pm$ one standard deviation over these models. Training-trajectory rollouts integrate Eq.~\eqref{eq:modal_ode} from the first observed modal state and use
\begin{align}
\label{eq:error_defs}
E_{\mathrm{rel}}(t_n):=\frac{\|U_{\mathrm{pred}}(t_n)-U_{\mathrm{true}}(t_n)\|_2}{\|U_{\mathrm{true}}(t_n)\|_2},
\qquad
\bar E_{\mathrm{roll}}:=\frac{1}{T}\sum_{n=1}^T E_{\mathrm{rel}}(t_n).
\end{align}
Here $U$ stacks all grid values of the evaluated fields. A rollout from a noisy observation includes both initial-state noise and coefficient error. For the Burgers, KdV, ZK, and decaying Navier--Stokes testing curves, $\bar E_{\rm test}$ denotes the same time-averaged relative error, first averaged over five fresh clean initial conditions and then over the three fitted noise seeds. The learned and true models start from the identical clean state in each test, so these curves exclude initial observation noise. 
Three considerations fix the held-out protocol. A held-out initial condition has to be a new sample of the regime the model was trained on, so its construction follows the training initial condition. For the deterministic systems it is a perturbation of the same low-mode state. For the turbulent systems it is a new draw from the same random ensemble. In Orszag--Tang the amplitude perturbation is limited to $\pm10\%$, which keeps the vortex pair and the current-sheet event it produces. A larger re-phased perturbation changes the transient, and we report it separately as a stress test. 

A deterministic transient is assessed by trajectory error, while a chaotic flow can only be assessed by invariant statistics, so the turbulent benchmarks use energy, flux, and spectrum errors in its place. These statistics have to be taken on the attractor. Each turbulent deployment therefore discards a spin-up at least as long as the time the true model needs to reach its stationary state, which is a few hundred time units for the laminar-to-turbulent transition of Kolmogorov flow, about one hundred for the linear growth and saturation of HW, and tens of $\gamma_{\max}^{-1}$ for mHW. The averaging window that follows spans several of the slowest time scales of each system. The true model integrated from the same initial conditions sets the finite-sampling floor against which the learned models are compared.
Deterministic held-out tests use five fresh initial conditions obtained by perturbing the training low-mode amplitudes within $\pm40\%$ and drawing new phases (the Orszag--Tang case additionally uses $\pm10\%$ perturbations without new phases). Chaotic and turbulent benchmarks are evaluated on invariant statistics (mean energy or particle flux, their relative errors, and the log-spectrum error) from five fresh initial conditions never used in training: perturbed low-mode states of the same $\pm40\%$ construction for Kolmogorov flow, and independent Gaussian random fields for HW and mHW. Each held-out deployment discards a spin-up ($500$, $150$, and $40\,\gamma_{\max}^{-1}$ time units for Kolmogorov flow, HW, and mHW) and averages over the following $200$, $150$, and $80\,\gamma_{\max}^{-1}$ units. The true model integrated from the same initial conditions sets the finite-sampling floor shown in each figure. Pointwise prediction panels for these benchmarks are training-trajectory rollouts from the noisy first snapshot. The Orszag--Tang trajectory and event diagnostics are defined in Sec.~\ref{sec:ot2d}; its pointwise panels use a new held-out initial condition. 

\section{Benchmarks and Results}\label{sec:experiments}

\subsection{Common settings}\label{sec:common_settings}
Table~\ref{tab:settings} collects, for the nine benchmark configurations, every governing equation, training initial condition, identification window, spatial state, sampling interval $\Delta t_{\rm s}$, total number of stored snapshots $T$, and rollout horizon. The HW system appears twice, with band-closed and with independent DNS data. Unless stated otherwise, experiments use uniform offset weights, adaptive Dormand--Prince integration throughout, and full-batch L-BFGS with at most 500 iterations and gradient tolerance $10^{-10}$. Noise follows Eq.~\eqref{eq:noise_model} with the subsets of $\eta\in\{0,0.01,0.02,0.05,0.10,0.20\}$ shown in the figures. Both the FD and WARP methods train on snapshots at the same sampling rate.\par

\begingroup
\footnotesize
\setlength{\tabcolsep}{2pt}
\renewcommand{\arraystretch}{1.13}
\begin{longtable}{@{}>{\raggedright\arraybackslash}p{0.115\linewidth}>{\raggedright\arraybackslash}p{0.28\linewidth}>{\raggedright\arraybackslash}p{0.26\linewidth}>{\raggedright\arraybackslash}p{0.21\linewidth}>{\centering\arraybackslash}p{0.070\linewidth}@{}}
\caption{Benchmark definitions and numerical protocols. All domains are periodic. Here $M$ is the largest retained Fourier mode per axis, $K$ is the basis size per field, $d=N_fK$ is the total modal-state dimension, $\Delta t_{\rm s}$ is the interval between stored snapshots, and $T$ is the total number of snapshots. The last column gives the rollout horizon $H$ in snapshot intervals. The abbreviations bi and fwd denote bidirectional and forward-only rollouts. Adaptive Dormand--Prince integration is used throughout.}\label{tab:settings}\\
\toprule
Benchmark & \makecell[l]{Governing equation and\\true coefficients} & \makecell[l]{Training initial condition and\\identification window} & \makecell[l]{Spatial state and\\time sampling} & \makecell{$H$\\(steps)} \\
\midrule
\endfirsthead
\multicolumn{5}{c}{\tablename\ \thetable\ continued}\\
\toprule
Benchmark & \makecell[l]{Governing equation and\\true coefficients} & \makecell[l]{Training initial condition and\\identification window} & \makecell[l]{Spatial state and\\time sampling} & \makecell{$H$\\(steps)} \\
\midrule
\endhead
\midrule
\multicolumn{5}{r}{Continued on next page}\\
\endfoot
\bottomrule
\endlastfoot

\rowcolor{gray!12}\multicolumn{5}{@{}l}{\textbf{Deterministic benchmarks}}\\
\hline
Burgers &
$u_t=buu_x+\nu u_{xx}$, $(b^\ast,\nu^\ast)=(-1,0.05)$ &
$u_0=\sin x+0.5\sin(2x+0.3)$ on $[0,2\pi]$, $t\in[0,0.4]$ &
64 grid points, $M=12$, $K=d=25$, $\Delta t_{\rm s}=10^{-3}$, $T=401$ & \makecell{$50$\\bi} \\
\hline
KdV &
$u_t=-\alpha uu_x-\beta u_{xxx}$, $(\alpha^\ast,\beta^\ast)=(6,1)$ &
$u_0=0.20\sin x+0.10\cos(2x-0.3)+0.06\sin(3x+0.7)+0.04\cos(5x-0.2)$, $t\in[0,0.1]$ &
128 grid points, $M=15$, $K=d=31$, $\Delta t_{\rm s}=10^{-5}$, $T=10{,}001$ & \makecell{$50$\\bi} \\
\hline
ZK &
$\begin{aligned}u_t={}&-\alpha uu_x\\&-\beta\partial_x\Delta u,\end{aligned}$\par $(\alpha^\ast,\beta^\ast)=(6,1)$ &
$u_0=0.20\sin x\cos y+0.10\cos(2x-0.3)\cos(y+0.5)+0.08\sin(x+0.4)\sin(2y-0.2)+0.05\cos(2x)\sin(2y)$, $t\in[0,0.016]$ &
$64^2$ grid, $M=4$, $K=d=81$, $\Delta t_{\rm s}=2\!\times\!10^{-5}$, $T=801$ & \makecell{$50$\\bi} \\
\hline
\makecell[l]{Decaying 2D\\vorticity} &
$\begin{aligned}\omega_t={}&-\alpha(u\!\cdot\!\nabla)\omega\\&+\nu\Delta\omega,\\ u={}&(\chi_y,-\chi_x),\quad \Delta\chi=-\omega,
\end{aligned}$\par $(\alpha^\ast,\nu^\ast)=(1,0.02)$ &
$\omega_0=\sin x\cos y+0.8\cos(x+0.4)\sin(y-0.2)+0.5\sin(2x-0.3)\cos(y+0.5)+0.4\cos x\sin2y$, $t\in[0,0.2]$ &
$64^2$ grid, $M=4$, $K=d=81$, $\Delta t_{\rm s}=5\!\times\!10^{-4}$, $T=401$ & \makecell{$50$\\bi} \\
\hline
\makecell[l]{Orszag--Tang\\MHD vortex} &
$\begin{aligned}\omega_t={}&-a_u(u\!\cdot\!\nabla)\omega\\&+a_B(B\!\cdot\!\nabla)j+\nu\Delta\omega,\\ A_t={}&-a_A(u\!\cdot\!\nabla)A+\eta_B\Delta A
\end{aligned}$\par\smallskip $\begin{aligned}(a_u^\ast,a_B^\ast,\nu^\ast)&=(1,1,0.025),\\(a_A^\ast,\eta_B^\ast)&=(1,0.025)
\end{aligned}$ &
$\chi_0=2(\cos x+\cos y)$ and $A_0=2\cos x+\cos2y$ on $[0,2\pi]^2$, $t\in[0.15,1.5]$ &
$64^2$ grid, $M=21$, $K=1{,}849$, $d=3{,}698$, $\Delta t_{\rm s}=0.005$, $T=271$ & \makecell{$200$\\fwd} \\

\hline
\rowcolor{gray!12}\multicolumn{5}{@{}l}{\textbf{Chaotic and turbulent benchmarks}}\\
\hline
\makecell[l]{Kolmogorov\\flow} &
$\begin{aligned}\omega_t={}&-a(u\!\cdot\!\nabla)\omega+\nu\Delta\omega\\&+A_f f(y),\\ f(y)={}&-k_f\cos(k_fy),
\end{aligned}$\par $(a^\ast,\nu^\ast,A_f^\ast)=(1,1/40,1)$ &
Laminar state plus a fixed $10^{-2}$ low-mode perturbation on $[0,2\pi]^2$, $t\in[500,600]$ after spin-up &
$64^2$ grid, $M=21$, $K=d=1{,}849$, $\Delta t_{\rm s}=0.05$, $T=2{,}001$ & \makecell{$10$\\fwd} \\
\hline
HW &
$\begin{aligned}n_t={}&-\{\phi,n\}+c_1(\phi-n)\\&-\kappa\phi_y-D_n(-\Delta)^pn,\\ \omega_t={}&-\{\phi,\omega\}+c_1(\phi-n)\\&-D_w(-\Delta)^p\omega,\\ \omega={}&\Delta\phi
\end{aligned}$\par\smallskip $c_1\in\{0.1,1,5\}$ and $\kappa^\ast=1$ &
Independent mean-free Gaussian random fields for $n$ and $\omega$ with covariance $0.01I$. Saturated windows are $[250,350]$ for $c_1\le1$ and $[1250,1350]$ for $c_1=5$ &
Fourier band $M=12$, $K=625$, $d=1{,}250$, $\Delta t_{\rm s}=0.01$, $T=10{,}001$ & \makecell{$50$ fwd\\$+\,5$ bwd} \\
\hline
HW DNS &
HW equation above with $c_1\in\{0.1,1,5\}$ and $\kappa^\ast=1$. Here $D_n^\ast=D_w^\ast=5\!\times\!10^{-9}$ &
Independent \texttt{hw2d} DNS sampled in saturation. Windows are $[200,300]$, $[250,350]$, and $[1250,1350]$ for increasing $c_1$ &
$512^2$ DNS grid projected to $M\in\{12,16,21\}$, $K\le1{,}849$, $d\le3{,}698$, $\Delta t_{\rm s}=0.05$, $T=2{,}001$ & \makecell{$10$\\fwd} \\
\hline
mHW &
$\begin{aligned}n_t={}&-\{\phi,n\}+C(\tilde\phi-\tilde n)\\&-\kappa\phi_y-D(-\Delta)^2n,\\ \omega_t={}&-\{\phi,\omega\}+C(\tilde\phi-\tilde n)\\&-\nu(-\Delta)^2\omega,\\ \omega={}&\Delta\phi
\end{aligned}$\par\smallskip $C\in\{0.01,0.1,0.3,1\}$, $\kappa^\ast=1$, and $D^\ast=\nu^\ast$ &
Independent spectrally enveloped Gaussian random fields for $\phi$ and $n$. The window is $\gamma_{\max}t\in[60,100]$ in a box holding ten fastest-growing wavelengths &
Fourier band $M=21$, $K=1{,}849$, $d=3{,}698$, $\gamma_{\max}\Delta t_{\rm s}=0.05$, $T=801$ & \makecell{$10$\\fwd} \\
\end{longtable}

\par\endgroup
\subsection{Burgers' equation}
\label{sec:burgers_experiments}

Burgers' equation~\citep{Bateman1915,Burgers1948} is the simplest nonlinear benchmark in Table~\ref{tab:settings}, combining quadratic convection and diffusion. It isolates the competition between nonlinear steepening and viscous smoothing before the coupled fluid and plasma examples.

\begin{figure}
\begin{minipage}[t]{0.55\linewidth}
\centering
\includegraphics[width=\linewidth]{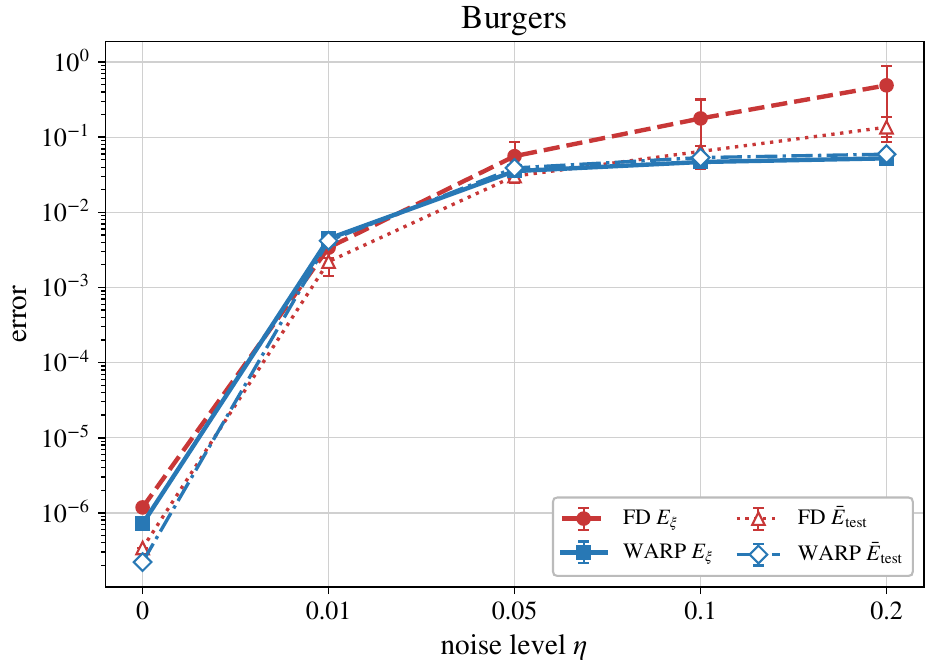}
\caption{Coefficient error $E_\xi$ (filled markers) and testing rollout error $\bar E_{\rm test}$ (open markers) versus observation-noise level for Burgers. Red: FD; blue: WARP. Testing uses five fresh clean initial conditions, with learned and true models launched from the same state. Error bars show one standard deviation over three noise seeds after averaging the testing error over initial conditions}
\label{fig:burgers_noise}
\end{minipage}\hfill
\begin{minipage}[t]{0.43\linewidth}
\centering
\includegraphics[width=\linewidth]{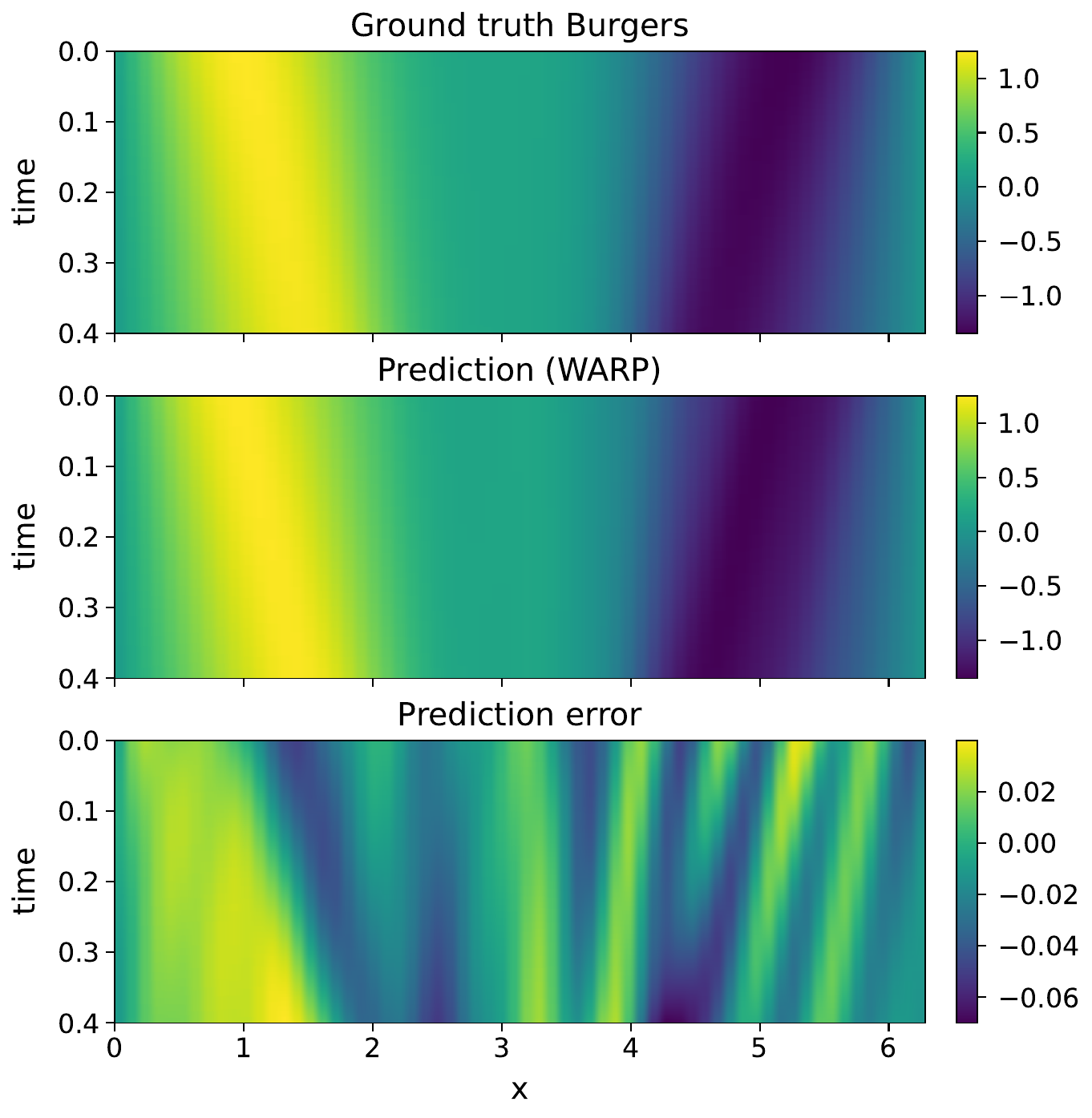}
\caption{Burgers at $5\%$ noise: WARP rollout from the observed initial state}
\label{fig:burgers_physical_rollout}
\end{minipage}
\end{figure}

Figure~\ref{fig:burgers_noise} compares coefficient error and testing rollout error from fresh initial conditions across noise levels. On noise-free data both methods recover the coefficients essentially exactly. At $1\%$ noise they remain statistically comparable because the Burgers library is the best-conditioned of the benchmarks. The slight WARP excess at this point is associated with the diffusion coefficient, which both estimators underestimate and WARP attenuates slightly more on this benchmark. Such attenuation is consistent with errors-in-variables bias; weak projection leaves noise in both the response and the candidate library~\citep{BortzWENDy2023,MessengerBortzConsistency2025}. These references explain the bias mechanism; they do not establish a universal direction or magnitude of bias for WARP. The WARP coefficient advantage then grows steadily with noise and reaches about an order of magnitude at the highest level. 
At the highest noise level the FD convection estimate becomes strongly seed-dependent. WARP recovers it to within a few percent with a seed-to-seed spread more than an order of magnitude smaller. Burgers is the one benchmark on which FD generalizes better at low noise ($0.01\le\eta\le0.05$), consistent with the attenuated diffusion coefficient. Figure~\ref{fig:burgers_physical_rollout} shows a representative physical-space rollout at $5\%$ noise.

\subsection{Korteweg--de Vries equation}
\label{sec:kdv_experiments}

The Korteweg--de Vries (KdV) equation, first derived for shallow-water waves~\citep{KortewegDeVries1895} and later for ion-acoustic solitons~\citep{WashimiTaniuti1966}, adds the third-order dispersive operator shown in Table~\ref{tab:settings}. Pointwise estimation of this derivative amplifies noise as $O(\Delta x^{-3})$~\citep{Chartrand2011,vanBreugel2020}. In the weak modal form no spatial derivative of the data is taken, but the dispersive library column scales as $k^3$, so the noise in the highest retained modes enters the regression with the largest weight, and the central time difference at $\Delta t_{\rm s}=10^{-5}$ amplifies the snapshot noise; the rollout objective replaces that difference by integrated increments over many snapshots (Appendix~\ref{sec:mechanism}). We also tested the modified KdV equation with a cubic nonlinear term and found very similar relative performance, so that case is not shown.

\begin{figure}
\begin{minipage}[t]{0.55\linewidth}
\centering
\includegraphics[width=\linewidth]{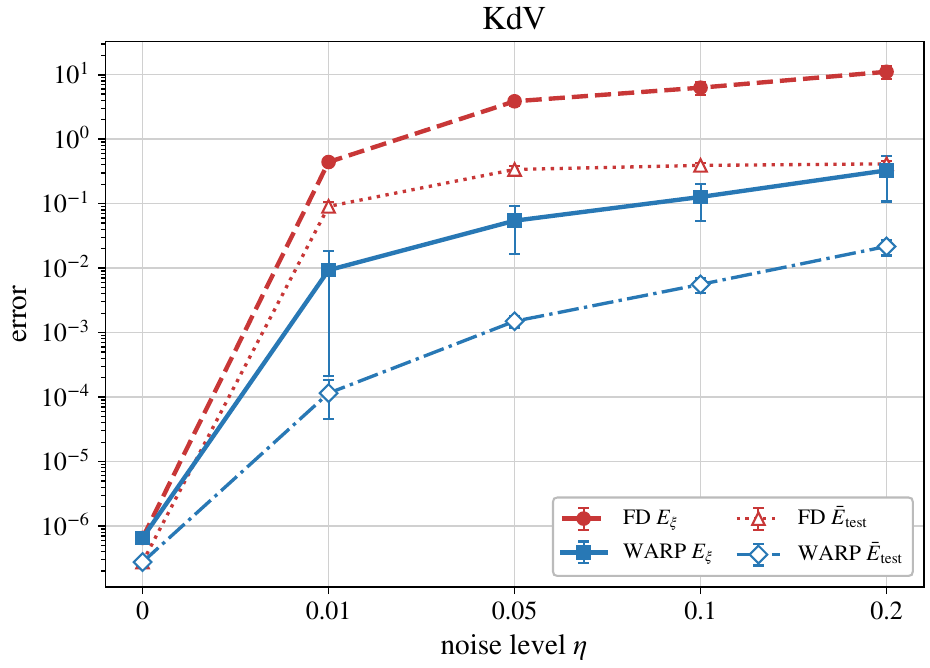}
\caption{KdV coefficient error $E_\xi$ and testing rollout error $\bar E_{\rm test}$, with the same five-initial-condition protocol and marker/error-bar conventions as Fig.~\ref{fig:burgers_noise}}
\label{fig:kdv_noise}
\end{minipage}\hfill
\begin{minipage}[t]{0.43\linewidth}
\centering
\includegraphics[width=\linewidth]{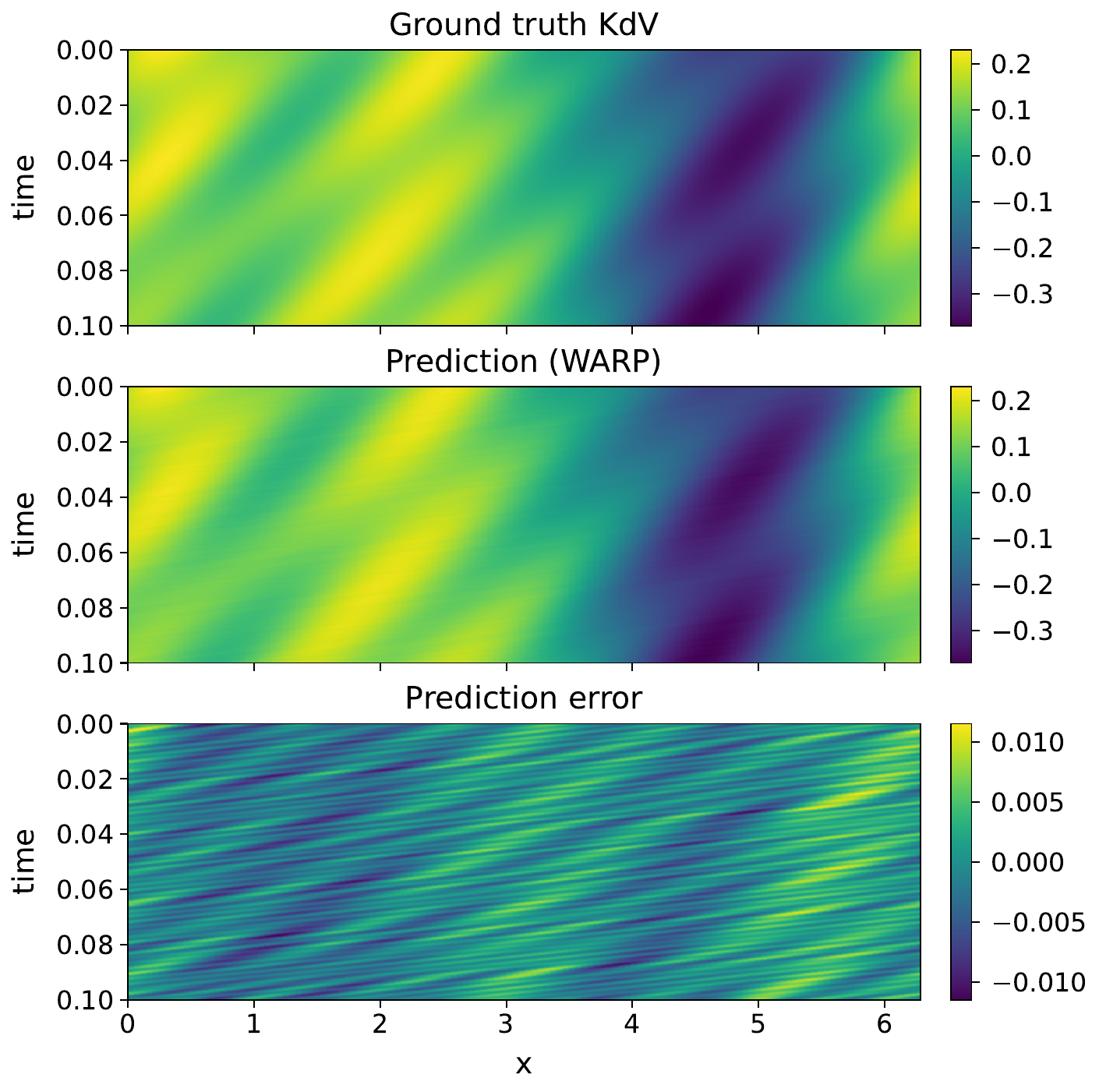}
\caption{KdV at $5\%$ noise, panels as in Fig.~\ref{fig:burgers_physical_rollout}}
\label{fig:kdv_rollout}
\end{minipage}
\end{figure}

Figure~\ref{fig:kdv_noise} shows that KdV is substantially more noise-sensitive for FD, which loses the dispersive coefficient almost entirely at the highest noise level, while WARP recovers both coefficients throughout. The coefficient improvement is one to two orders of magnitude. In the same plot, testing rollout error from fresh initial conditions is one to three orders of magnitude smaller for WARP at every nonzero noise level. Figure~\ref{fig:kdv_rollout} shows the trajectory, autonomous rollout, and pointwise error at $5\%$ noise.

\subsection{Zakharov--Kuznetsov equation (2D)}
\label{sec:zk_experiments}

The Zakharov--Kuznetsov (ZK) equation is the canonical 2D generalization of KdV, originally derived for weakly nonlinear ion-acoustic waves in a magnetized plasma~\citep{ZakharovKuznetsov1974}. Its anisotropic dispersive operator is listed in Table~\ref{tab:settings} and reduces to KdV when $\partial_y=0$.

\begin{figure}
\begin{minipage}[t]{0.55\linewidth}
\centering
\includegraphics[width=\linewidth]{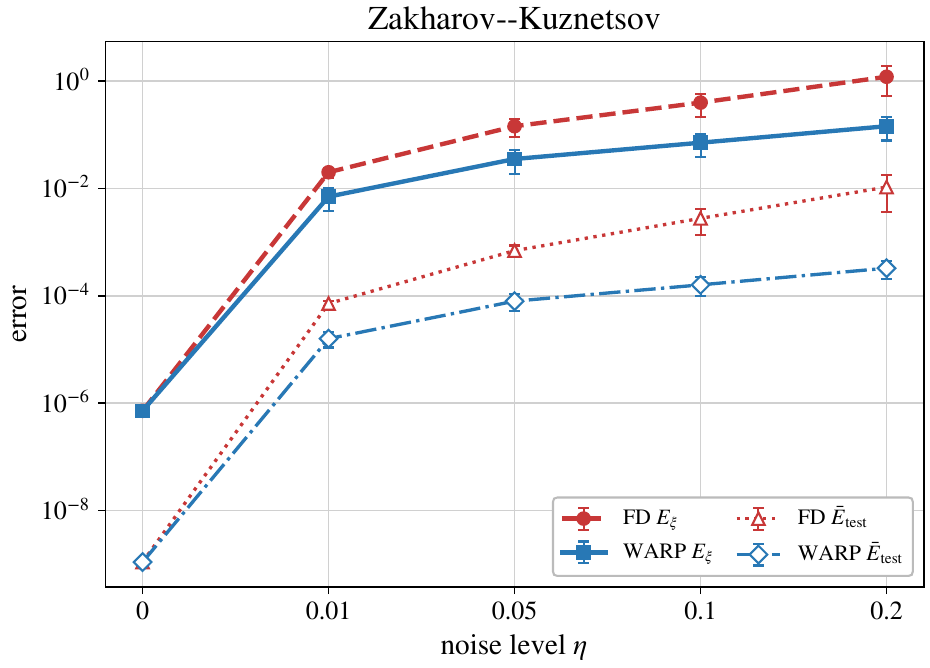}
\caption{2D Zakharov--Kuznetsov coefficient error $E_\xi$ and testing rollout error $\bar E_{\rm test}$, with the same five-initial-condition protocol and marker/error-bar conventions as Fig.~\ref{fig:burgers_noise}}
\label{fig:zk_noise}
\end{minipage}\hfill
\begin{minipage}[t]{0.43\linewidth}
\centering
\includegraphics[width=\linewidth]{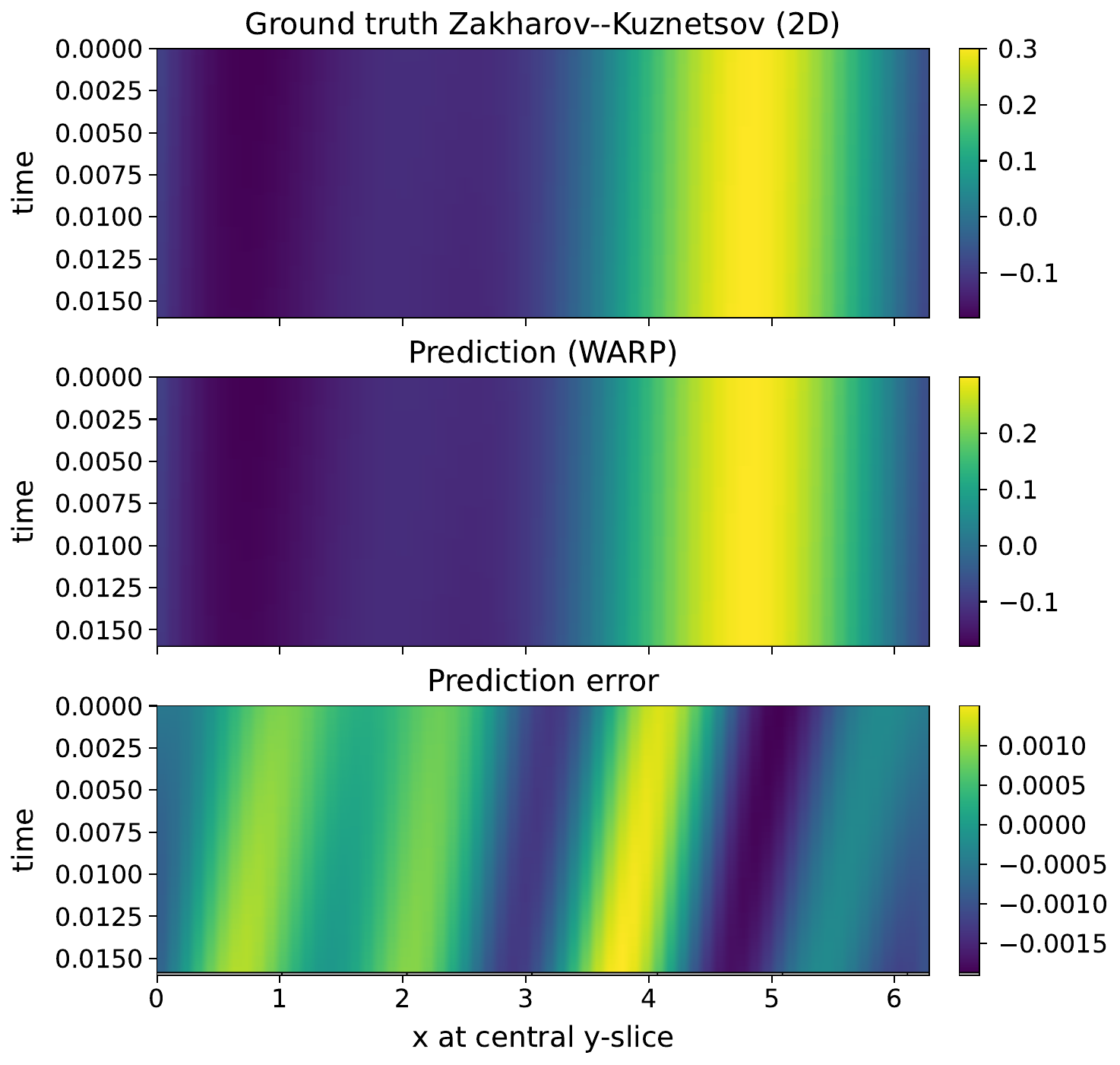}
\caption{ZK at $5\%$ noise along the central $y$-slice, panels as in Fig.~\ref{fig:burgers_physical_rollout}. The nonzero error at $t=0$ is the observation noise of the initial snapshot projected onto the retained band}
\label{fig:zk_rollout}
\end{minipage}
\end{figure}

Figure~\ref{fig:zk_noise} reports coefficient and testing rollout errors. Under noise, WARP's coefficient advantage grows monotonically to roughly an order of magnitude. The testing curves start both models from the same clean initial conditions and show the resulting improvement in integrated predictions. Figure~\ref{fig:zk_rollout} shows the trajectory, autonomous rollout, and pointwise error at $5\%$ noise along the central $y$-slice.

\subsection{Decaying 2D incompressible Navier--Stokes vorticity}
\label{sec:ns2d_experiments}

The incompressible Navier--Stokes vorticity equation introduces genuine fluid advection that is quadratic and nonlocal in vorticity through the streamfunction relation in Table~\ref{tab:settings}. This case tests recovery of a velocity-coupled nonlinear operator from a decaying 2D flow.

\begin{figure}
\begin{minipage}[t]{0.55\linewidth}
\centering
\includegraphics[width=\linewidth]{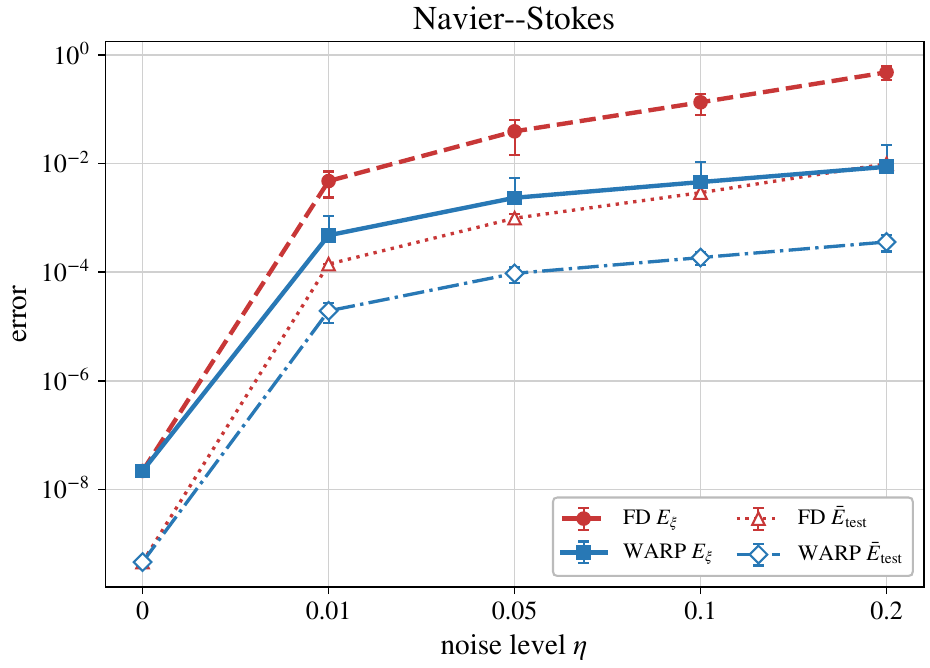}
\caption{2D Navier--Stokes vorticity coefficient error $E_\xi$ and testing rollout error $\bar E_{\rm test}$, with the same five-initial-condition protocol and marker/error-bar conventions as Fig.~\ref{fig:burgers_noise}}
\label{fig:ns2d_noise}
\end{minipage}\hfill
\begin{minipage}[t]{0.43\linewidth}
\centering
\includegraphics[width=\linewidth]{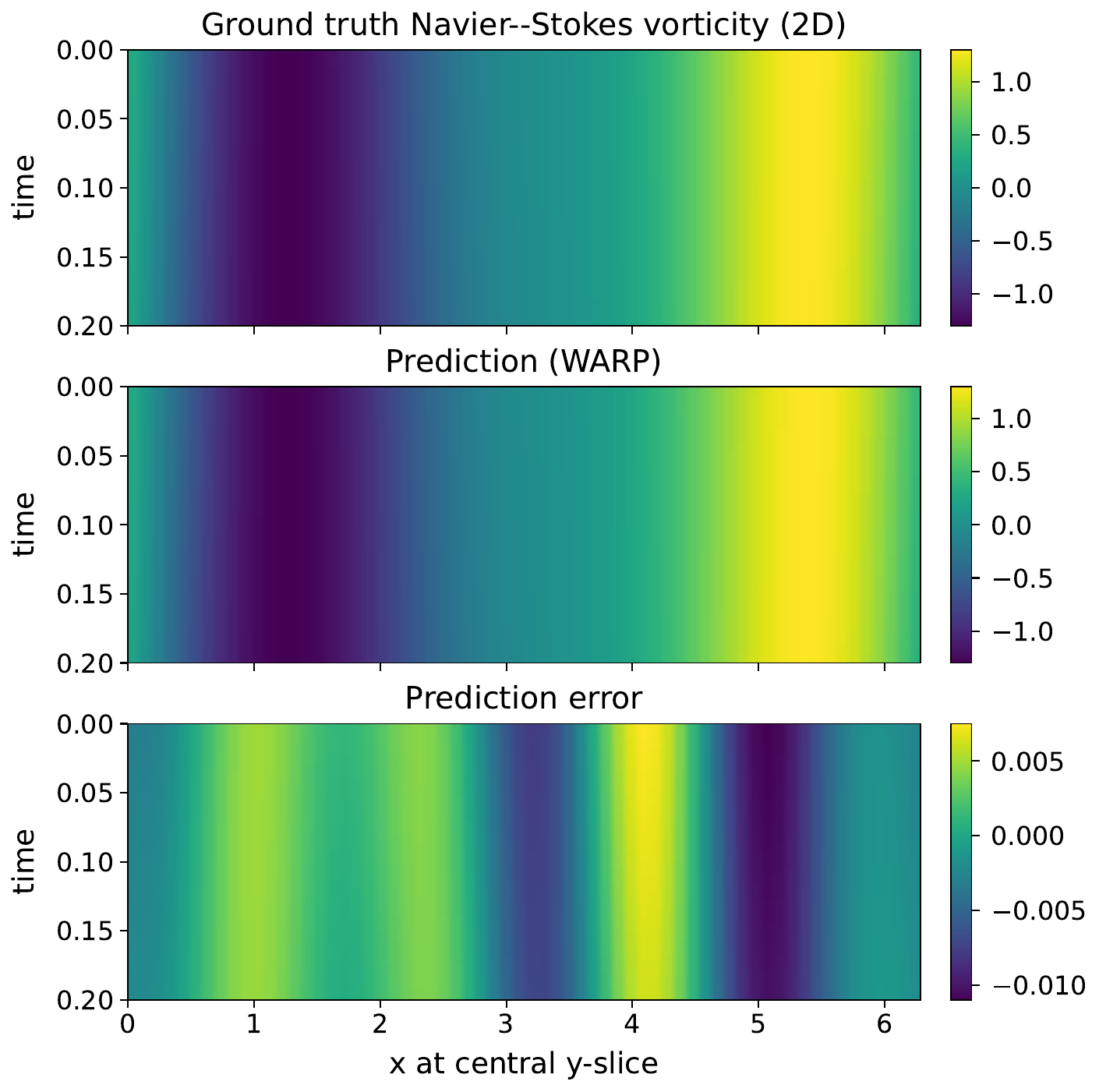}
\caption{Navier--Stokes vorticity at $5\%$ noise along the central $y$-slice, with panels as in Fig.~\ref{fig:burgers_physical_rollout}. The nonzero error at $t=0$ has the origin described for Fig.~\ref{fig:zk_rollout}}
\label{fig:ns2d_rollout}
\end{minipage}
\end{figure}

Figure~\ref{fig:ns2d_noise} reports a coefficient advantage that grows from one to nearly two orders of magnitude with noise; at the highest level FD underestimates the advection coefficient by about half while WARP remains within a percent. As for ZK, testing rollouts from identical clean initial conditions reveal the prediction improvement without a contribution from initial observation noise. Figure~\ref{fig:ns2d_rollout} shows the trajectory, autonomous rollout, and pointwise error at $5\%$ noise along the central $y$-slice.

\subsection{Body-forced Kolmogorov turbulence (2D)}
\label{sec:kolm}

Kolmogorov flow adds monochromatic body forcing to sustain statistically stationary 2D turbulence~\citep{MeshalkinSinai1961,ChandlerKerswell2013}. It is the first benchmark with a state-independent forcing term. Weak-form discovery has also been demonstrated on simulated and laboratory flows of this class~\citep{Reinbold2020,Reinbold2021}; here, the identified coefficients must additionally sustain the flow under autonomous integration.

\begin{figure}
\centering
\includegraphics[width=\linewidth, keepaspectratio]{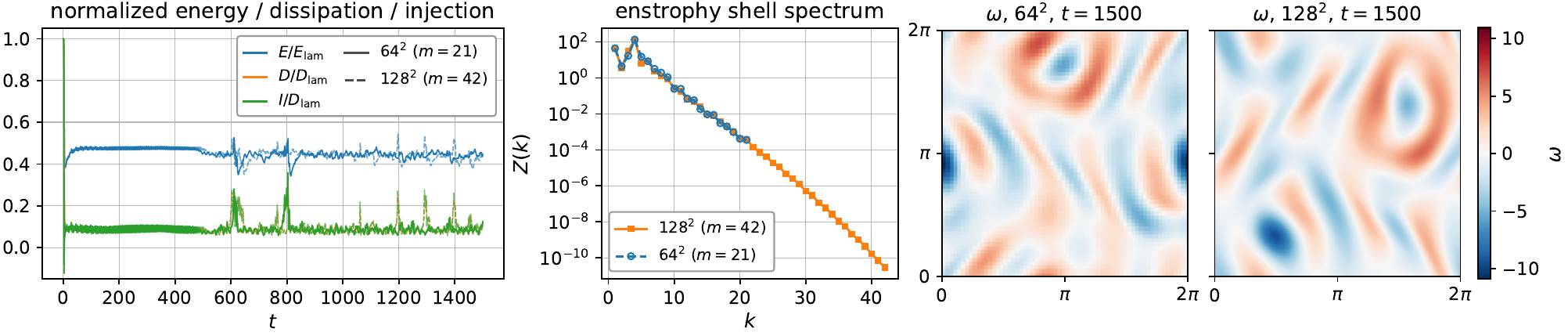}
\caption{Solver validation for Kolmogorov flow at $64^2$ ($M=21$) and $128^2$ ($M=42$). The spectrum is taken at the end of the run; the two vorticity fields are different realizations of the chaotic state}
\label{fig:kolm_sanity}
\end{figure}

The discovery window $t\in[500,600]$ follows 500 time units of discarded spin-up, so identification uses the statistically stationary turbulent state and excludes the laminar-to-turbulent transient. This is the regime the benchmark is meant to test; the learned coefficients must sustain the saturated flow under autonomous integration, and held-out performance is evaluated using the same stationary statistics. Transient trajectory prediction is tested separately by the decaying benchmarks of Secs.~\ref{sec:burgers_experiments}--\ref{sec:ot2d}. The 100-unit discovery window contains several intermittent dissipation bursts and therefore samples the characteristic variability of the saturated state. The saturated energy and dissipation statistics reproduce the reference values, and a band-doubled companion run confirms resolution convergence (Fig.~\ref{fig:kolm_sanity}). The plotted diagnostics are the kinetic energy $E$, the viscous energy dissipation $D$, and the power $I$ injected by the body force, normalized by the laminar values $E_{\rm lam}$ and $D_{\rm lam}$; $Z(k)$ is the enstrophy in a wavenumber shell.

Training is forward-only with $H=10$. The adaptive training integrator subdivides each sampling interval as needed.

\begin{figure}
\begin{minipage}[t]{0.45\linewidth}
\centering
\includegraphics[width=\linewidth, keepaspectratio]{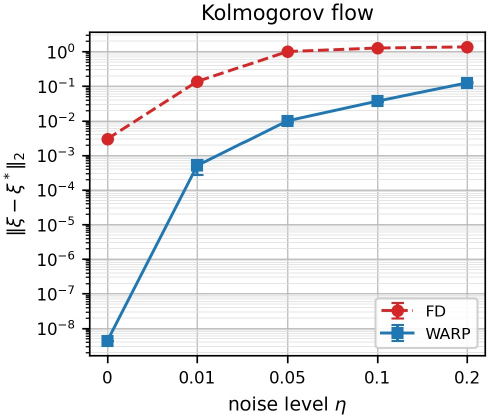}
\caption{Kolmogorov-flow coefficient error $\|\xi-\xi^\ast\|_2$ versus noise level, coefficient-error conventions as in Fig.~\ref{fig:burgers_noise}}
\label{fig:kolm_noise}
\end{minipage}\hfill
\begin{minipage}[t]{0.49\linewidth}
\centering
\includegraphics[width=0.9\linewidth, keepaspectratio]{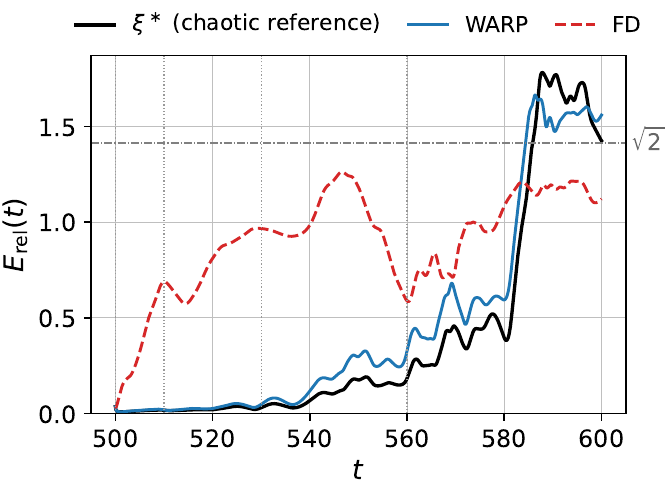}
\caption{Rollout error versus time for Kolmogorov flow at $\eta=5\%$ (seed 42): relative $L_2$ error of the vorticity state against the clean reference for WARP, FD, and the true coefficients $\xi^\ast$, all deployed with adaptive Dormand--Prince integration from the same noisy first snapshot}
\label{fig:kolm_errtime}
\end{minipage}
\end{figure}

Figure~\ref{fig:kolm_noise} reports coefficient recovery. At zero noise the central-difference estimate has an $O(\Delta t_{\rm s}^2)$ truncation bias set by the sampling rate. Under noise, its coefficient error approaches unity as all three coefficients attenuate together. WARP degrades gradually and preserves the coefficient signs and magnitudes. The same separation holds when the library is enlarged to all terms up to quadratic in the fields and second order in derivatives, without a sparsity penalty (Appendix~\ref{app:biglib}).

\begin{figure}
\centering
\includegraphics[width=0.7\linewidth]{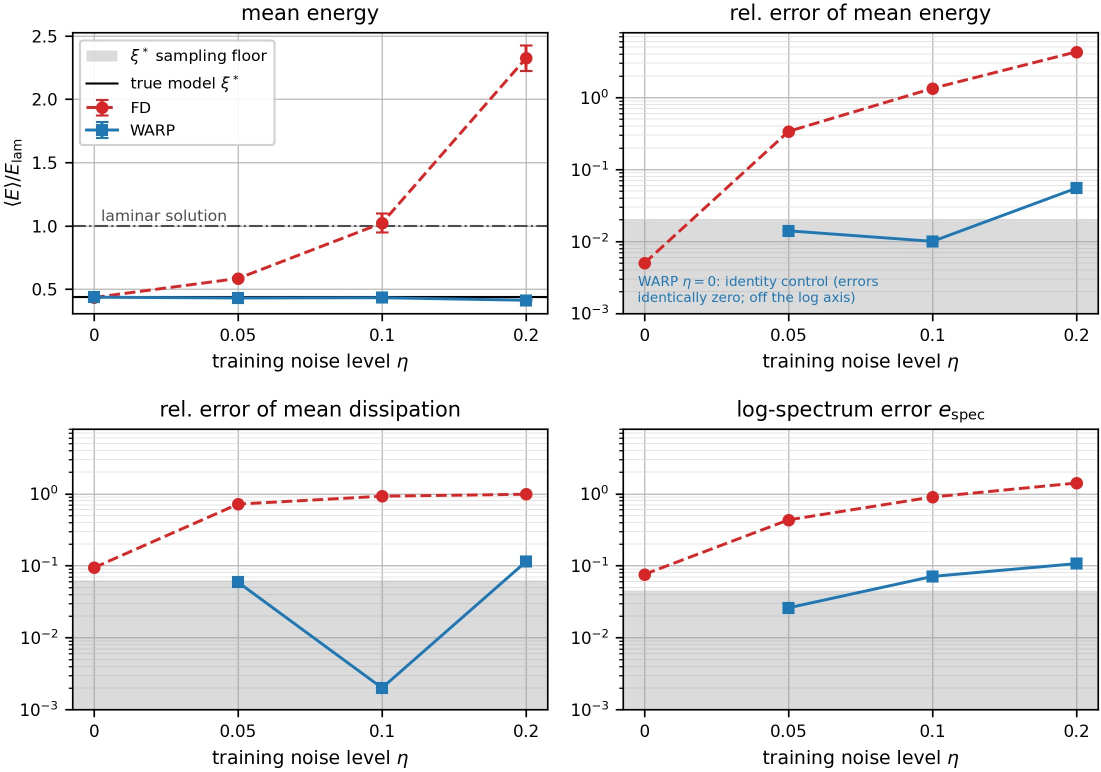}
\caption{Kolmogorov-flow statistics from five fresh initial conditions versus training noise}
\label{fig:kolm_stat_heldout}
\end{figure}

In deployments from fresh initial conditions, WARP remains statistically indistinguishable from the true model through $\eta=0.1$, while FD drifts past the laminar energy (Fig.~\ref{fig:kolm_stat_heldout}).

\begin{figure}
\centering
\includegraphics[width=\linewidth]{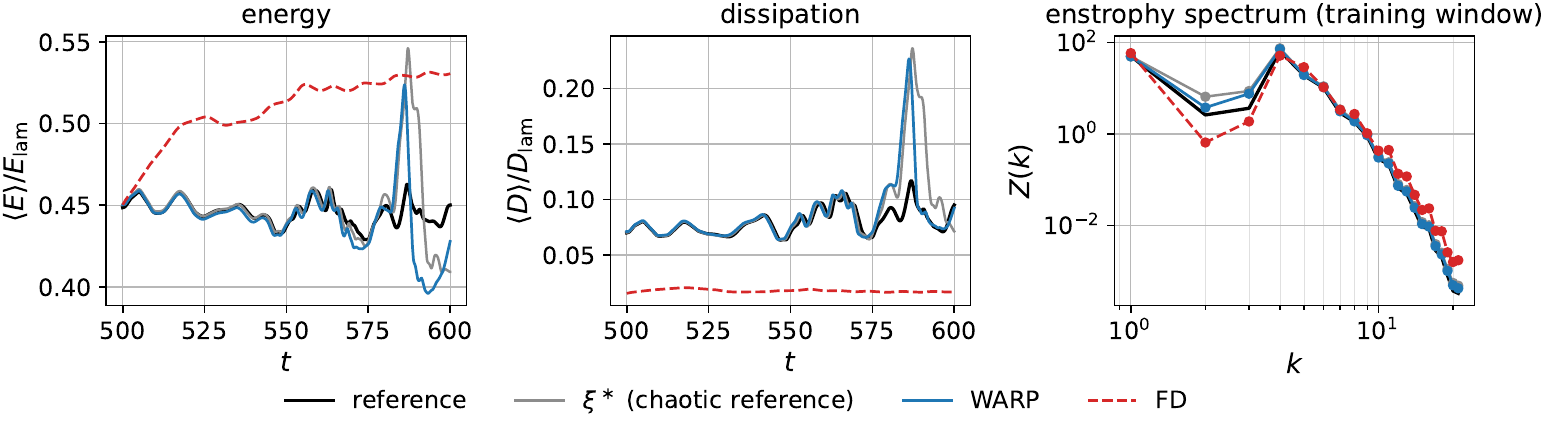}
\caption{Kolmogorov flow at $\eta=5\%$: energy and dissipation along the deployment of Fig.~\ref{fig:kolm_errtime}, and the enstrophy spectrum averaged over the discovery window. Black: clean trajectory. Grey: true coefficients $\xi^\ast$ restarted from the same noisy first snapshot as WARP and FD}
\label{fig:kolm_deploy_diag}
\end{figure}

\begin{figure}
\centering
\includegraphics[width=\linewidth]{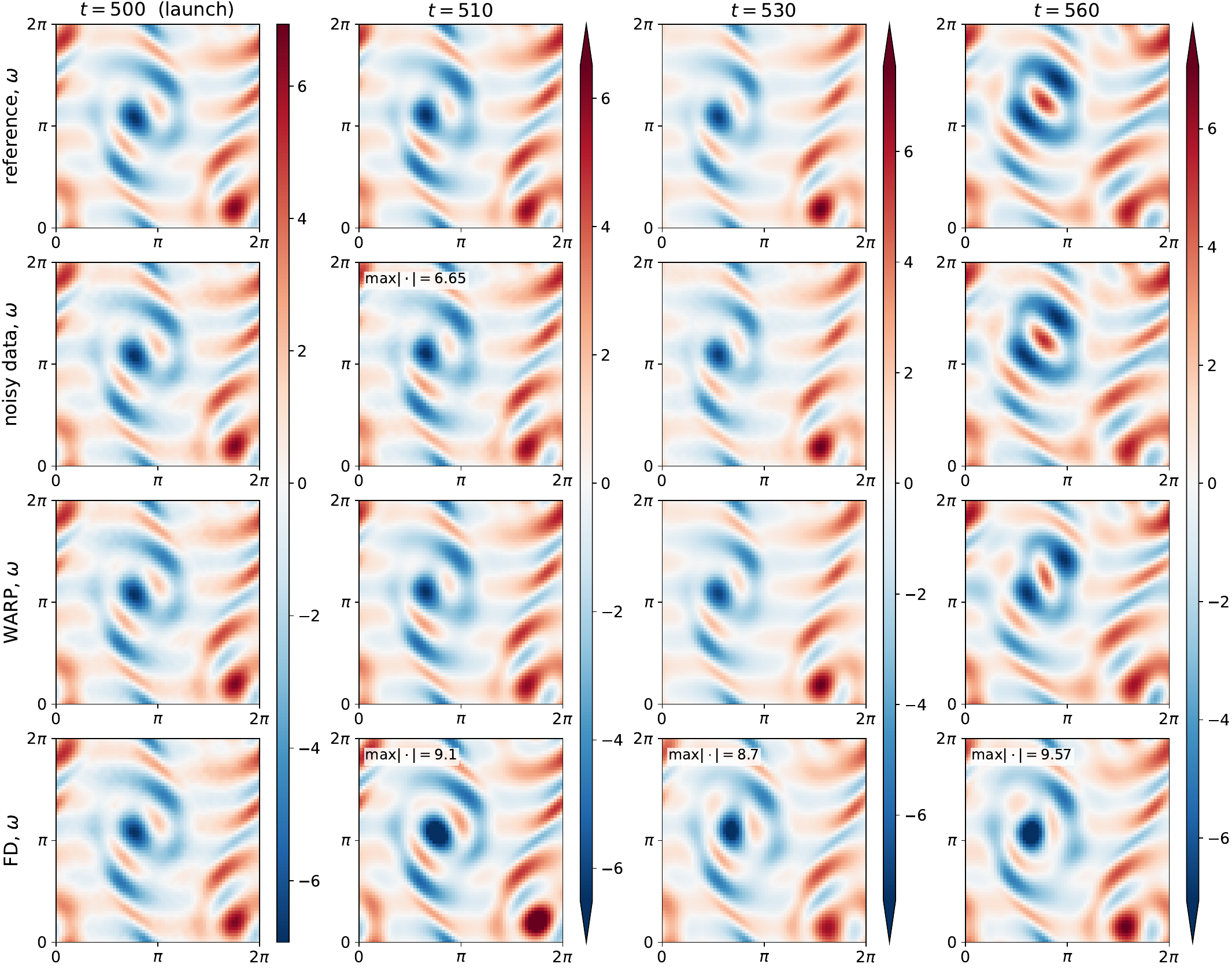}
\caption{
Kolmogorov-flow vorticity at $\eta=5\%$ (seed 42), with WARP and FD launched from the noisy snapshot at $t=500$. Color scales follow the reference in each column; panels that exceed the range are annotated with their peak}
\label{fig:kolm_fields}
\end{figure}

Figures~\ref{fig:kolm_errtime} and~\ref{fig:kolm_fields} show the deployment at $\eta=5\%$, launched from the noisy first snapshot of the discovery window. Two trajectories use the true coefficients $\xi^\ast$: the black curve is the clean data-generating trajectory, while the grey curve is a new integration restarted from the noisy first snapshot. Their separation therefore measures amplification of the initial observation error under the true dynamics. WARP (blue) and FD (red) start from that same noisy snapshot, so comparison with grey measures the additional effect of learned-coefficient error. WARP follows this reference curve over more than one hundred training horizons and remains pointwise indistinguishable from the reference through the middle columns, while FD's error reaches order one within ten time units and remains below the decorrelation level thereafter, so its flow stays partially correlated with the reference even though its vortices are too intense; the attenuated viscosity removes too little enstrophy, consistent with the dissipation deficit in Fig.~\ref{fig:kolm_stat_heldout}.
Figure~\ref{fig:kolm_deploy_diag} shows the same deployments on the diagnostics used to validate the data. The WARP energy, dissipation, and enstrophy spectrum follow the true coefficients through a chaotic burst, whereas FD drifts to a higher energy, loses most of the dissipation, and shifts enstrophy from the large scales to the top of the band.

\subsection{Hasegawa--Wakatani coupled-field discovery in saturated turbulence}
\label{sec:hw_turb}

\begin{figure}
\centering
\includegraphics[width=\linewidth, keepaspectratio]{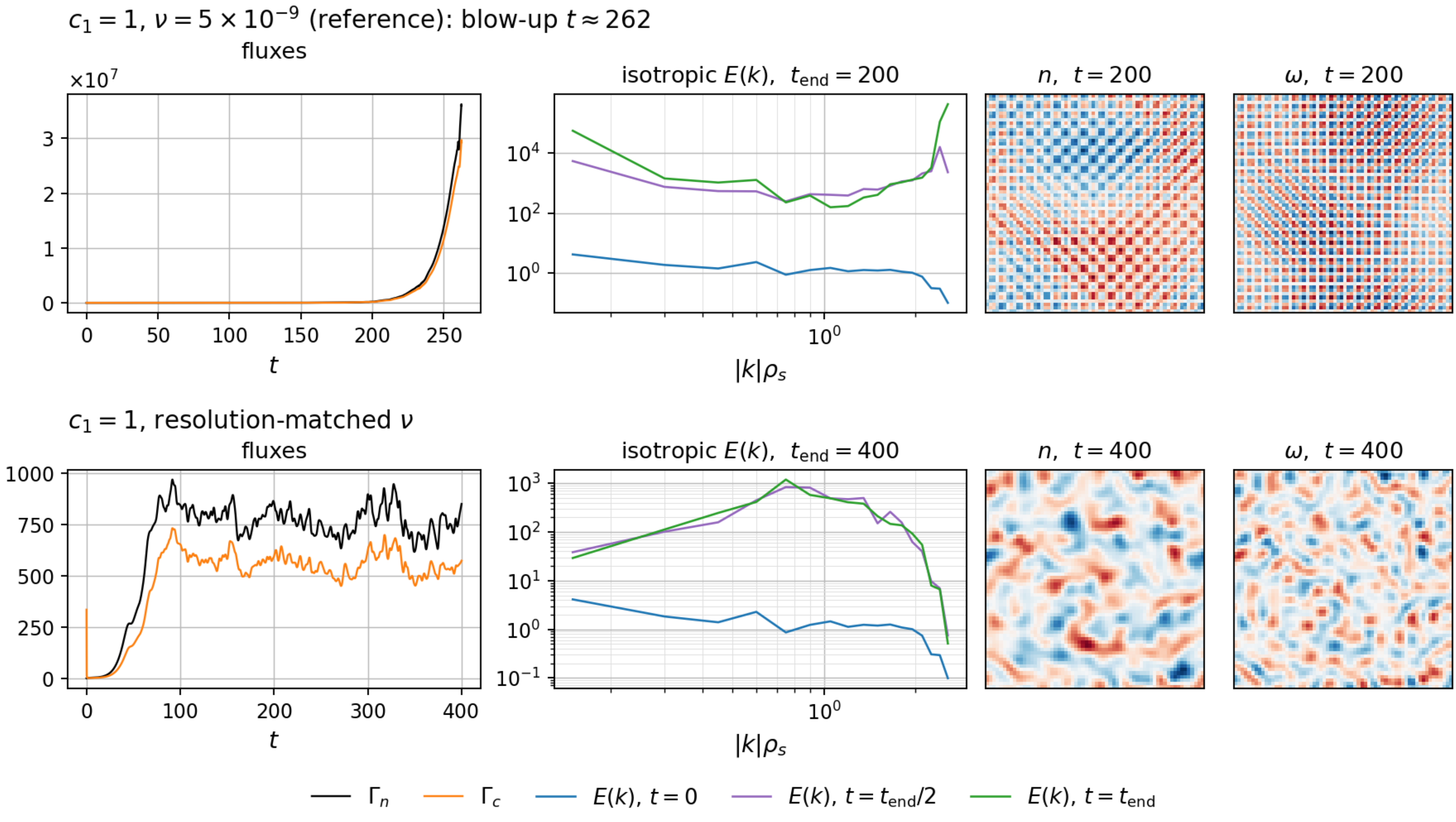}
\caption{Saturated HW data-generation diagnostics at $c_1=1$ on the $M=12$ band, with DNS-scale (top) and resolution-matched (bottom) hyperdiffusion. Other $c_1$: Appendix~\ref{app:hw_additional}}
\label{fig:hw_turb_sanity}
\end{figure}

The 2D Hasegawa--Wakatani (HW) system~\citep{PhysRevLett} models resistive drift-wave dynamics at the tokamak edge and evolves density fluctuation $n$ together with the vorticity $\omega=\Delta\phi$ of the $E\times B$ flow. Its shared advection, resistive coupling, density-gradient drive, and dissipation give seven independently learned coefficients. We use the ordinary HW system in the saturated configuration of \citet{GahrFarcasJenko2024}, with a resolution-matched dissipation that sustains multivortex turbulence in the closed $M=12$ band (Fig.~\ref{fig:hw_turb_sanity}). The DNS-scale hyperdiffusion $\nu=5\times10^{-9}$ provides an insufficient small-scale sink on this truncated band, and that run blows up (Fig.~\ref{fig:hw_turb_sanity}, top).

The plotted diagnostics are the radial particle flux $\Gamma_n=-\int_\Omega n\,\partial_y\phi\,dx\,dy$, resistive energy dissipation $\Gamma_c=c_1\int_\Omega(\phi-n)^2\,dx\,dy$, and total fluctuation energy $E=\tfrac12\int_\Omega(n^2+|\nabla\phi|^2)\,dx\,dy$~\citep{GahrFarcasJenko2024}. Division by $L^2$ gives the corresponding per-area quantities. In the spectral panels $E(k)$ is energy summed over a wavenumber shell; $|k|\rho_s$ is wavenumber normalized by the ion-sound gyroradius $\rho_s$. The labels $n$ and $\omega$ denote density fluctuation and vorticity, and $t_{\rm end}$ is the final recorded time.

\begin{figure}
\begin{minipage}[t]{0.41\linewidth}
\centering
\includegraphics[width=0.95\linewidth, keepaspectratio]{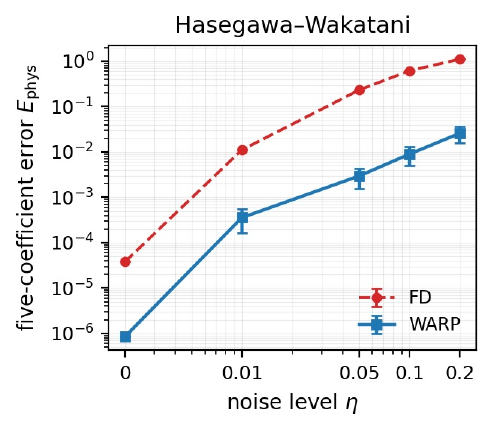}
\caption{Saturated HW coefficient error $E_\xi$ at $c_1=1$ versus noise level, coefficient-error conventions as in Fig.~\ref{fig:burgers_noise}. Training uses $H=50$ forward and five backward snapshot intervals. Error bars show $\pm1$ standard deviation over three noise seeds; the $\eta=0$ point is deterministic}
\label{fig:hw_turb_noise}
\end{minipage}\hfill
\begin{minipage}[t]{0.49\linewidth}
\centering
\includegraphics[width=0.95\linewidth, keepaspectratio]{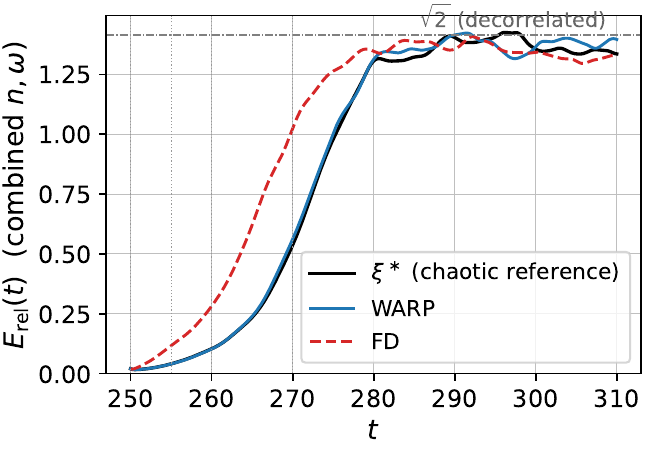}
\caption{Rollout error versus time for saturated HW turbulence at $c_1=1$ and $\eta=5\%$ for seed 42. Curves show relative $L_2$ error of the combined $(n,\omega)$ state against the clean reference for WARP, FD, and the true coefficients $\xi^\ast$, all deployed from the same noisy first snapshot. The $\xi^\ast$ curve isolates chaotic decorrelation; dotted lines mark the times shown in Fig.~\ref{fig:hw_turb_rollout}}
\label{fig:hw_turb_errtime}
\end{minipage}
\end{figure}

For the HW and projected-DNS plots we report $E_\xi$ over all seven coefficients. The learned hyperdiffusion coefficients are at most $10^{-2}$ in absolute value, since the band model absorbs part of the unresolved dissipation, and contribute less than $0.1\%$ to $E_\xi$. Under observation noise, WARP improves $E_\xi$ by one to two orders of magnitude at $c_1=1$ (Fig.~\ref{fig:hw_turb_noise}).

\begin{figure}
\centering
\includegraphics[width=\linewidth]{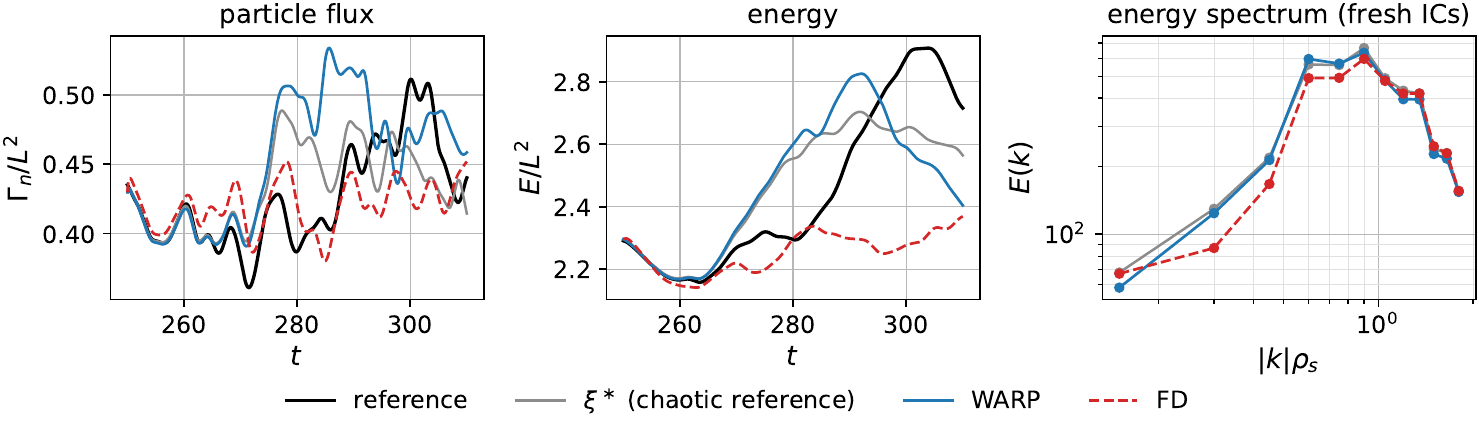}
\caption{Saturated HW at $c_1=1$ and $\eta=5\%$: particle flux and energy along the deployment of Fig.~\ref{fig:hw_turb_errtime}, and the energy spectrum averaged over the fresh initial conditions of Fig.~\ref{fig:hw_stat_heldout}. In the time traces, black is the clean trajectory and grey is the true model restarted from the noisy first snapshot; in the spectrum, grey is the true-model average}
\label{fig:hw_deploy_diag}
\end{figure}

Figures~\ref{fig:hw_turb_rollout} and~\ref{fig:hw_turb_errtime} show the deployment at $\eta=5\%$, launched from the noisy first snapshot. The WARP rollout remains visually indistinguishable from the reference in both fields through $t=260$, twenty training horizons, and its error curve coincides with that of the true coefficients deployed from the same noisy snapshot. As in the Kolmogorov case, black denotes the clean reference and grey denotes the true model restarted from the noisy observation; their separation measures sensitivity to the initial perturbation.
Agreement with the true-coefficient baseline indicates that the subsequent error growth is dominated by chaotic sensitivity to the initial perturbation. The FD model's error is several times larger during the growth phase and reaches saturation noticeably earlier.
Figure~\ref{fig:hw_deploy_diag} shows the same deployments on the flux, energy, and spectrum. After decorrelation the traces are distinct chaotic realizations, so only their level and fluctuation amplitude are comparable: the WARP traces fluctuate at the reference level, whereas FD runs a persistently low energy and a depleted large-scale spectrum.

\begin{figure}
\centering
\includegraphics[width=0.95\linewidth]{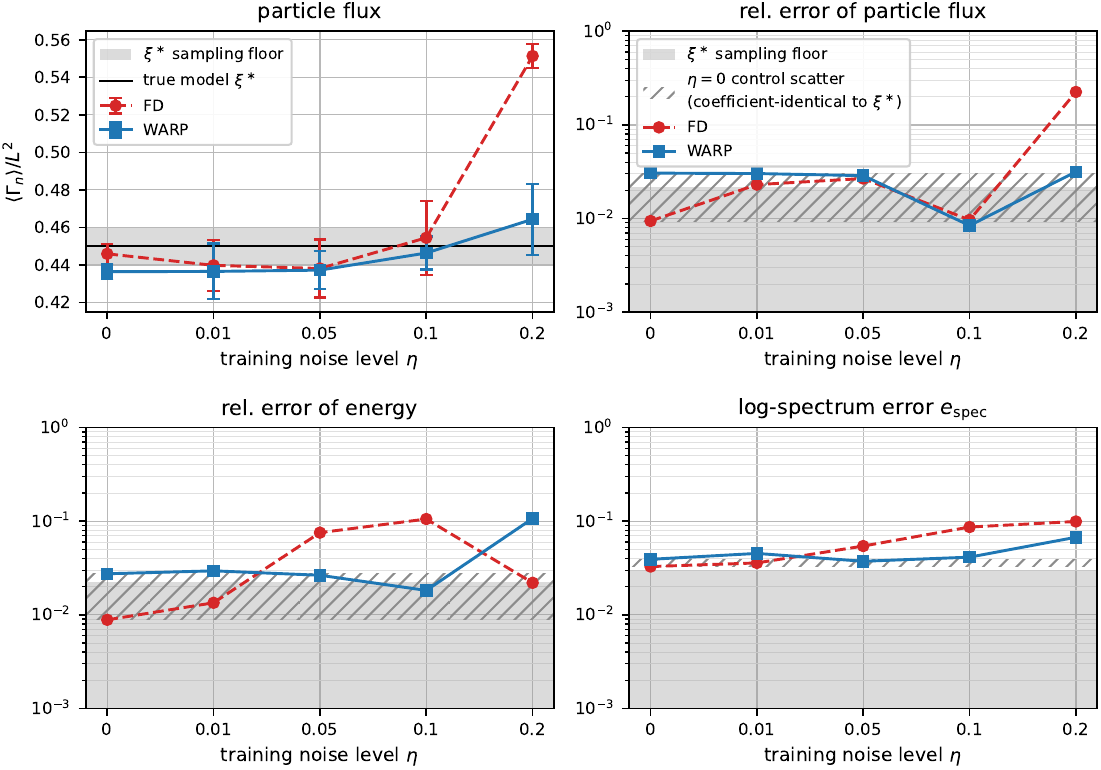}
\caption{Saturated HW at $c_1=1$: statistics from five fresh initial conditions versus training noise. Error bars show one standard deviation over the initial conditions}
\label{fig:hw_stat_heldout}
\end{figure}

From five fresh initial conditions, WARP remains within the scatter of the nearly coefficient-identical $\eta=0$ controls through $\eta=10\%$ and deviates only mildly at the highest noise level, where its energy error exceeds FD's while its flux and spectrum errors remain smaller, whereas FD separates in energy and spectrum from $5\%$ noise and overestimates the particle flux at high noise as the attenuated density coupling drives the model toward the high-flux hydrodynamic limit (Fig.~\ref{fig:hw_stat_heldout}). These tests use band-closed data. The independent-DNS experiment in Sec.~\ref{sec:hw_dns512} separately tests training data produced by a different spatial discretization.

\subsection{Closure discovery from independent \texorpdfstring{$512^2$}{512-squared} DNS data}
\label{sec:hw_dns512}

We next drop the closed-band assumption. Independent \texttt{hw2d} data~\citep{Greif2023hw2d} use the parameters of \citet{GahrFarcasJenko2024}, a $512^2$ grid, $\nu=5\times10^{-9}$, $p=3$, Arakawa brackets~\citep{Arakawa1966}, finite-difference spatial operators, and adaptive Dormand--Prince integration. The fields are sampled every $0.05$ time unit. The DNS reproduces the published flux ranges at $c_1=1$ (Fig.~\ref{fig:dns512_traces}); the $c_1=0.1$ and $5$ validations are in Appendix~\ref{app:hw_additional}.

\begin{figure}
\centering
\includegraphics[width=\linewidth, keepaspectratio]{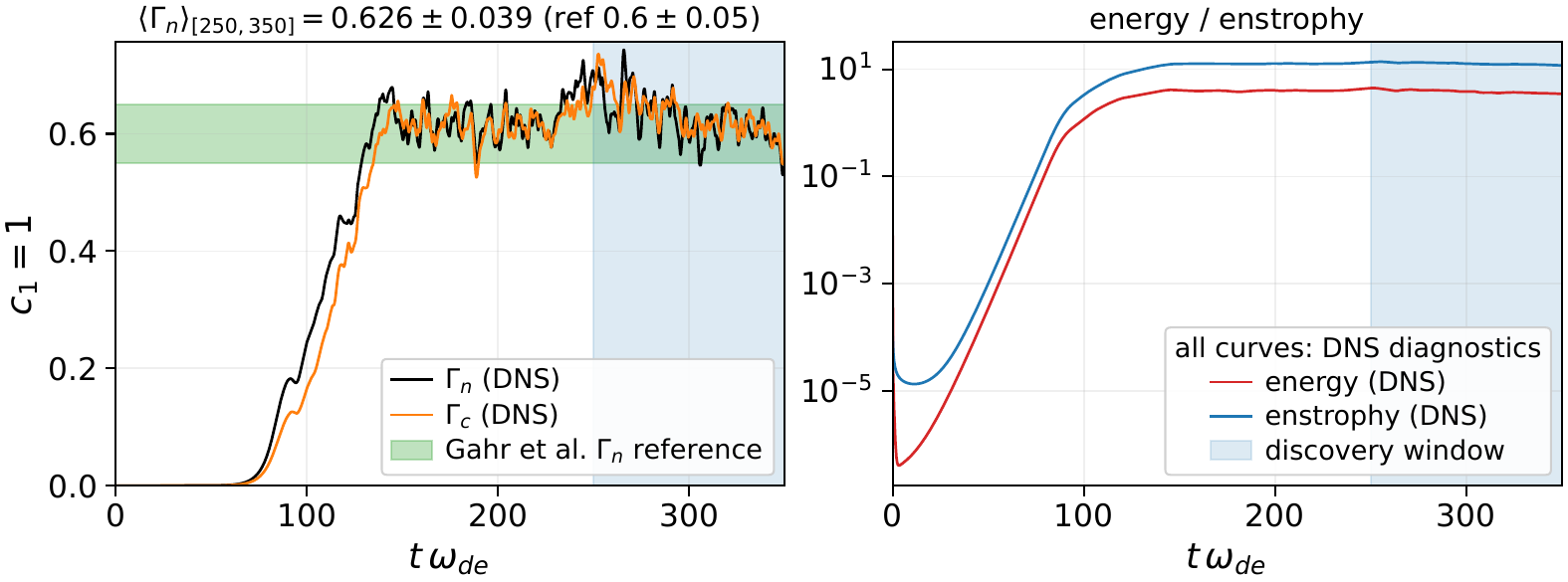}
\caption{Independent $512^2$ \texttt{hw2d} DNS at $c_1=1$, with the reference flux range of \citet{GahrFarcasJenko2024}. Other $c_1$: Appendix~\ref{app:hw_additional}}
\label{fig:dns512_traces}
\end{figure}

The projected band exchanges energy with discarded scales, so the retained operator library need not reproduce its dynamics exactly for any coefficient vector. We refer to the nonzero coefficient error on noise-free projected data as a closure-induced offset; it also includes differences between the data-generating and discovery discretizations. It is not a proven lower bound on estimation error. The truncated Poisson bracket lacks the subgrid stress that drains retained vorticity modes~\citep{Kraichnan1976}, and the data-generating finite-difference spatial discretization also differs from the spectral discovery model.

Figure~\ref{fig:dns512_closure} shows that WARP has lower coefficient error than FD at $c_1=1$ for each retained band, and that its zero-noise offset decreases as $M$ increases. Under observation noise, FD error rises while WARP remains near its zero-noise value. Appendix~\ref{app:hw_additional} shows the additional adiabaticities; WARP retains the same ordering there, with more pronounced noise sensitivity at $c_1=5$.

\subsection{Modified Hasegawa--Wakatani}
\label{sec:mhw}

The modified HW (mHW) system~\citep{NumataBallDewar2007} removes zonal components from the resistive coupling. Zonal modes then grow through Reynolds-stress transfer~\citep{DiamondZonalFlows2005}, and increasing adiabaticity carries the saturated state from eddy-dominated turbulence to zonal flow. We retain the seven-coefficient HW library and use full-field, resolution-matched fourth-order dissipation in a box containing ten fastest-growing wavelengths.

\begin{figure}
\begin{minipage}[t]{0.3\linewidth}
\centering
\includegraphics[width=0.95\linewidth]{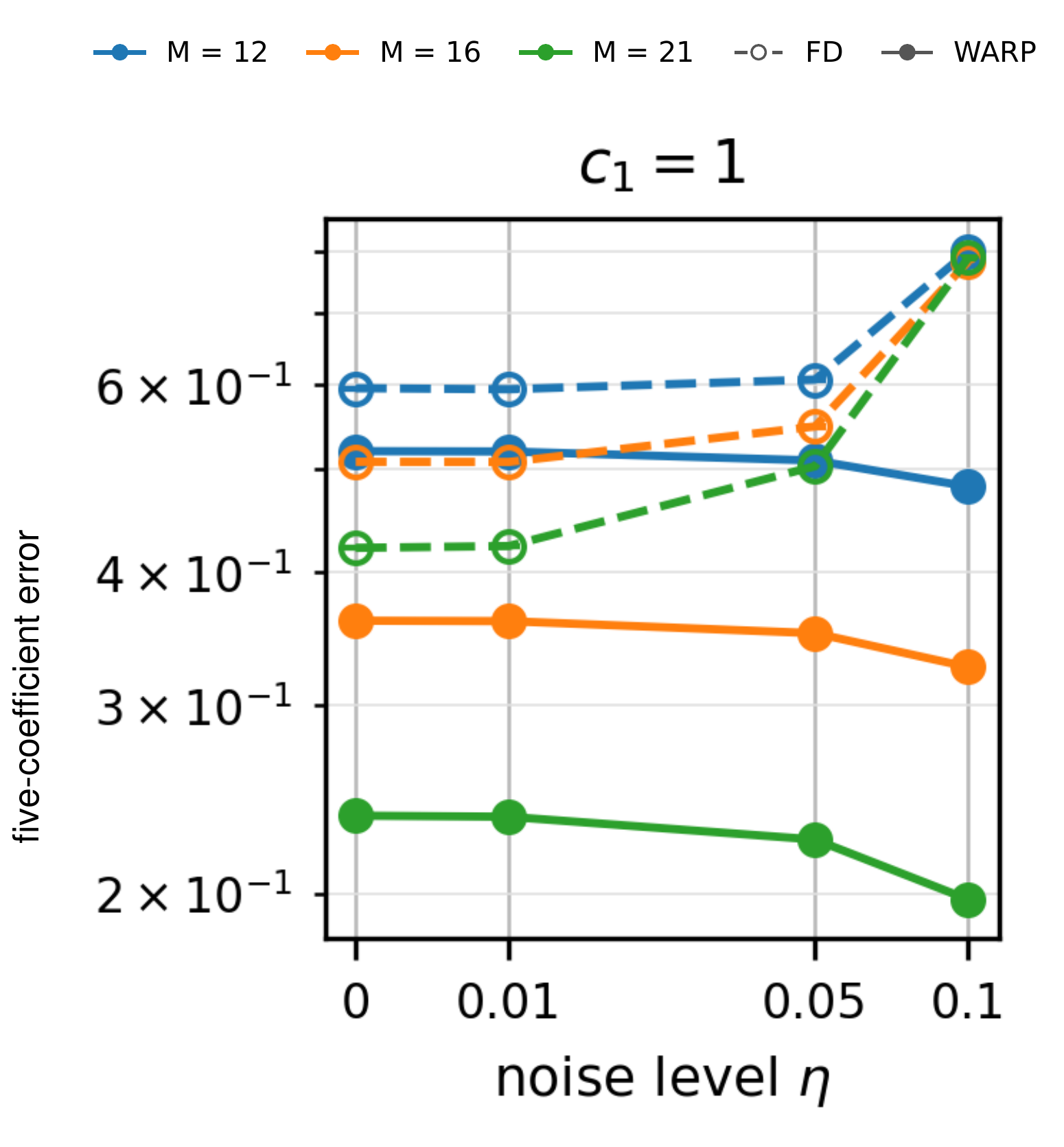}
\caption{Projected $512^2$ DNS at $c_1=1$: coefficient error $E_\xi$ versus noise level for three retained bands $M$. Other $c_1$ appear in Appendix~\ref{app:hw_additional}}
\label{fig:dns512_closure}
\end{minipage}\hfill
\begin{minipage}[t]{0.67\linewidth}
\centering
\includegraphics[width=0.99\linewidth,keepaspectratio]{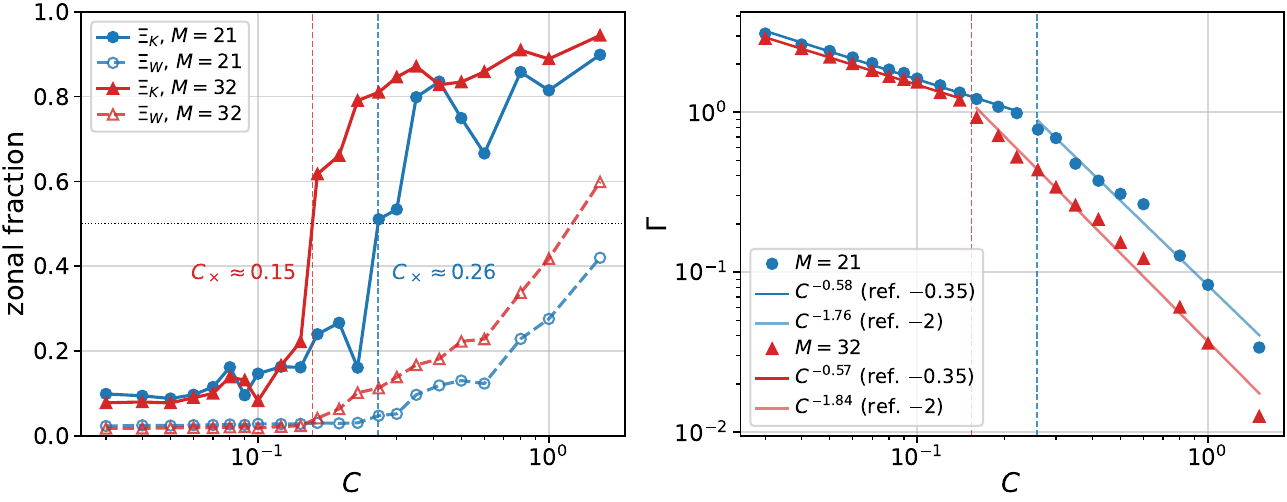}
\caption{Solver-only scan of the truncated mHW system at $M=21$ (circles) and $M=32$ (triangles) with resolution-matched $\nabla^4$ dissipation; each point is the final-quarter average of a single run. Left: zonal kinetic-energy fraction $\Xi_K$ (filled) and enstrophy fraction $\Xi_W$ (open) versus adiabaticity $C$; the dotted line marks $\Xi_K=\tfrac12$ and the dashed vertical lines the crossing $C_\times$ at each resolution. Right: particle flux $\Gamma$ versus $C$ with power-law fits below and above each crossing; the legend lists the fitted exponents together with the DNS exponents of \citet{GuillonGurcan2025}, which are fitted over much wider ranges of $C$}
\label{fig:mhw_transition}
\end{minipage}
\end{figure}

A solver-only scan over nineteen adiabaticities at two resolutions validates the truncated system against the reference phenomenology and fixes the identification cases (Fig.~\ref{fig:mhw_transition}). The zonal kinetic-energy fraction crosses one half within a narrow interval of $C$, and the particle flux follows two power laws below and above the crossing, as in the DNS of \citet{GuillonGurcan2025}. Doubling the band moves the crossing close to the DNS transition and brings the zonal-regime exponent within a tenth of the DNS value on the same window of $C$; the eddy-regime exponent is somewhat steeper than the DNS value and does not change with the band width, while the flux itself agrees with the DNS to within ten percent there. The $M=21$ scan therefore provides the two-regime system and located transition required by the identification experiment, which uses the truncated system as its own reference.
The four cases are $C\in\{0.01,\,0.1\}$ (eddy), $C=0.3$ (transition), and $C=1$ (zonal).

The peak linear growth rate sets the characteristic time scale. The discovery window is $\gamma_{\max}t\in[60,100]$ with Gaussian-random-field initial conditions, and training uses a ten-step forward rollout.
The eddy cases saturate into isotropic multivortex turbulence and the $C=1$ case into banded zonal flows, with the discovery windows inside the saturated phase.

\begin{figure}
\centering
\includegraphics[width=0.5\linewidth, keepaspectratio]{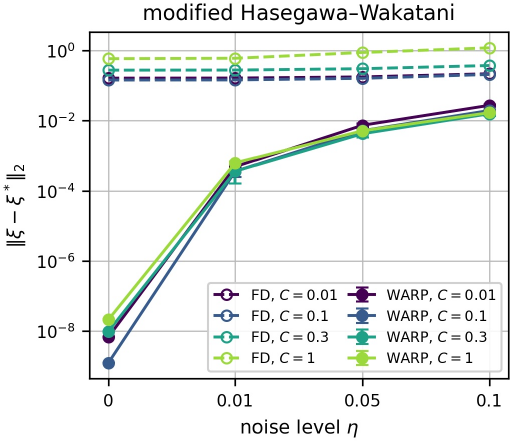}
\caption{mHW coefficient error $\|\xi-\xi^\ast\|_2$ versus noise level at four adiabaticities, with FD dashed and WARP solid}
\label{fig:mhw_noise}
\end{figure}

Figure~\ref{fig:mhw_noise} reports recovery across the four regimes. The FD curves sit on a sampling-rate bias floor, whereas WARP begins at the integration-accuracy floor and retains an advantage of roughly one to two orders of magnitude at $10\%$ noise.

\begin{figure}
\centering
\includegraphics[width=\linewidth,keepaspectratio]{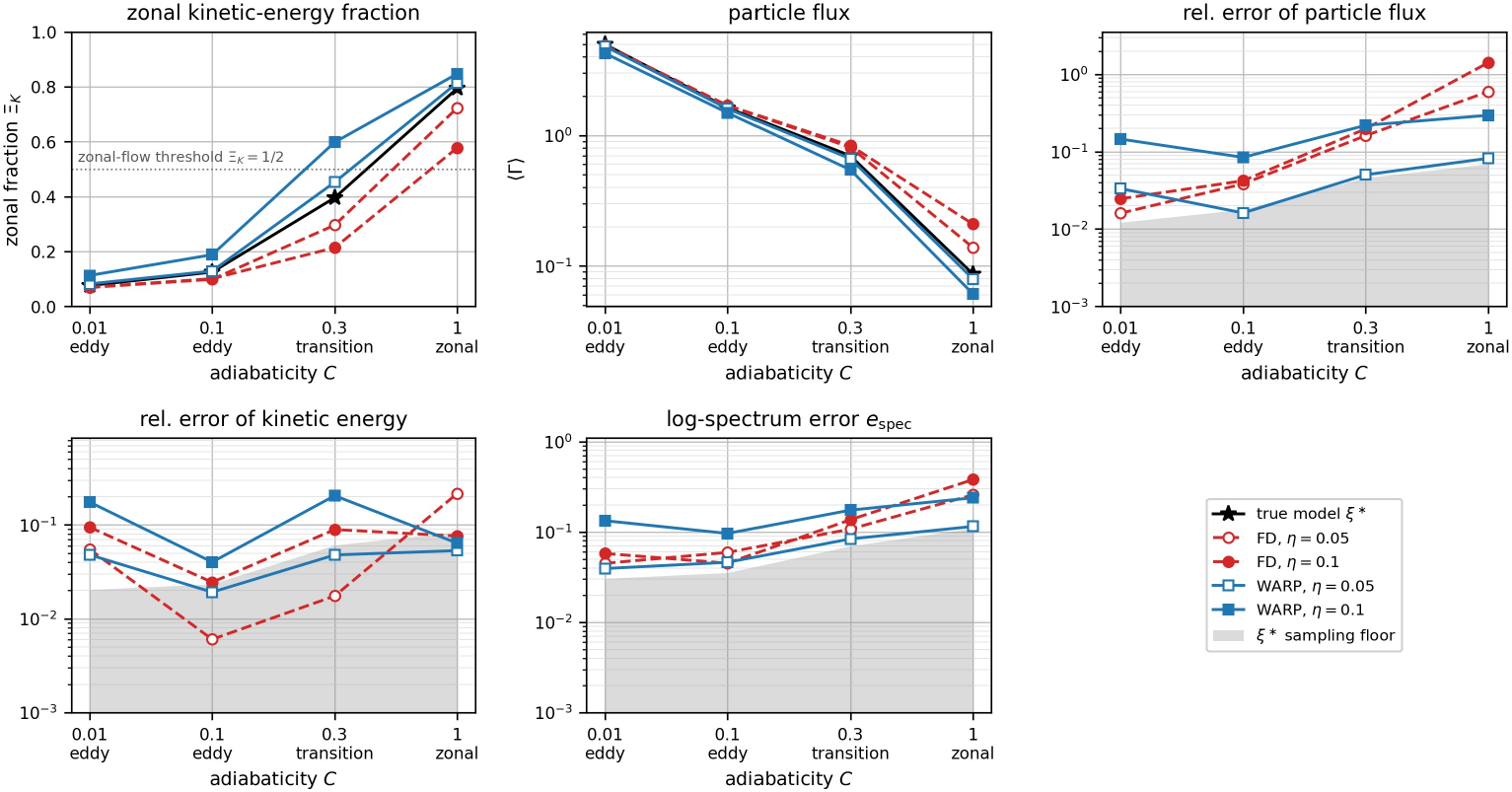}
\caption{mHW: statistics from five fresh initial conditions across the zonal-flow transition, for models trained at $\eta=0.05$ and $0.1$}
\label{fig:mhw_stat_heldout}
\end{figure}

From five fresh initial conditions, the WARP deployments at $5\%$ training noise are statistically indistinguishable from the true model for $C\ge0.1$, whereas FD separates most strongly in the zonal regime, where its flux error grows to order one while the zonal fraction drifts toward the eddy side (Fig.~\ref{fig:mhw_stat_heldout}). At $10\%$ noise the residual WARP coefficient error shifts the zonal fraction upward in every regime; near the transition this is enough to carry it across its threshold, and in the eddy regime it costs more in flux and energy than FD's uniformly attenuated coefficients, whose downward bias the eddy state tolerates. Invariant statistics are therefore not a monotone function of coefficient error, and the zonal regime is where the two estimators differ decisively.

\begin{figure}
\centering
\begin{subfigure}[t]{0.48\linewidth}
\centering
\includegraphics[width=\linewidth,keepaspectratio]{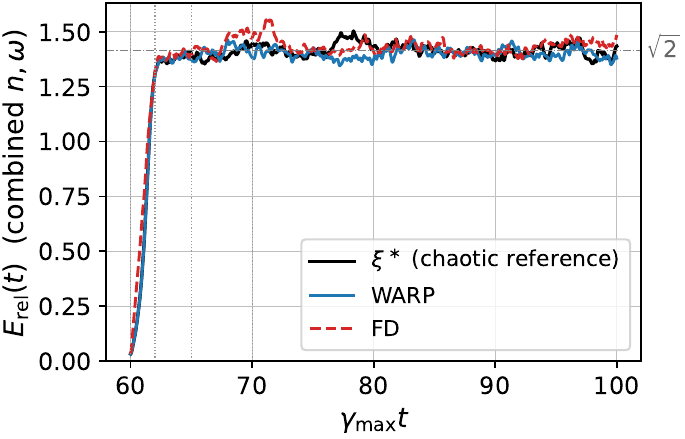}
\caption{$C=0.1$ (eddy)}
\end{subfigure}\hfill
\begin{subfigure}[t]{0.48\linewidth}
\centering
\includegraphics[width=\linewidth,keepaspectratio]{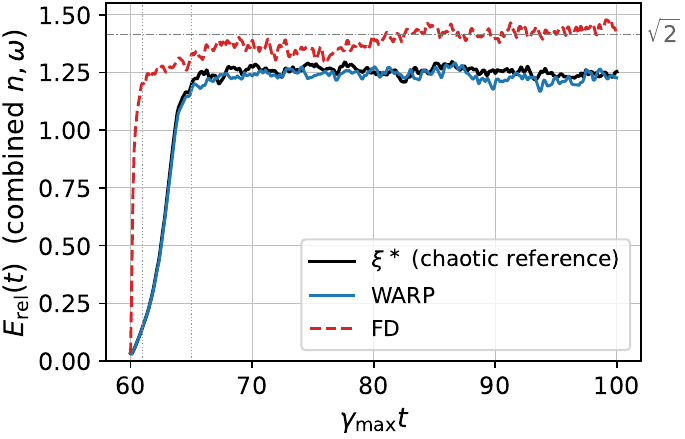}
\caption{$C=1$ (zonal)}
\end{subfigure}
\caption{mHW at $\eta=5\%$ (seed 42): rollout error against the clean reference, with all three models launched from the same noisy first snapshot at $\gamma_{\max}t=60$. Dotted lines mark one and five growth times after launch}
\label{fig:mhw_errtime}
\end{figure}

Figure~\ref{fig:mhw_errtime} compares prediction errors when all three models start from the same noisy observation. In the eddy-dominated case ($C=0.1$), even the true coefficients lose pointwise agreement with the clean trajectory within a few instability growth times. Most of the later error therefore reflects chaotic sensitivity to the initial perturbation, so we assess those predictions using the statistical tests in Fig.~\ref{fig:mhw_stat_heldout}. In the zonal-flow case ($C=1$), WARP follows the true-coefficient error curve, whereas FD departs earlier and reaches a larger error. The slowly changing zonal component can remain correlated after the eddies decorrelate. Consequently, the total-state error can saturate below $\sqrt2$, the reference level for two decorrelated states of equal norm. A low pointwise error in this regime must therefore be read together with the particle-flux, zonal-energy, and spectrum diagnostics.

\begin{figure}
\centering
\includegraphics[width=\linewidth]{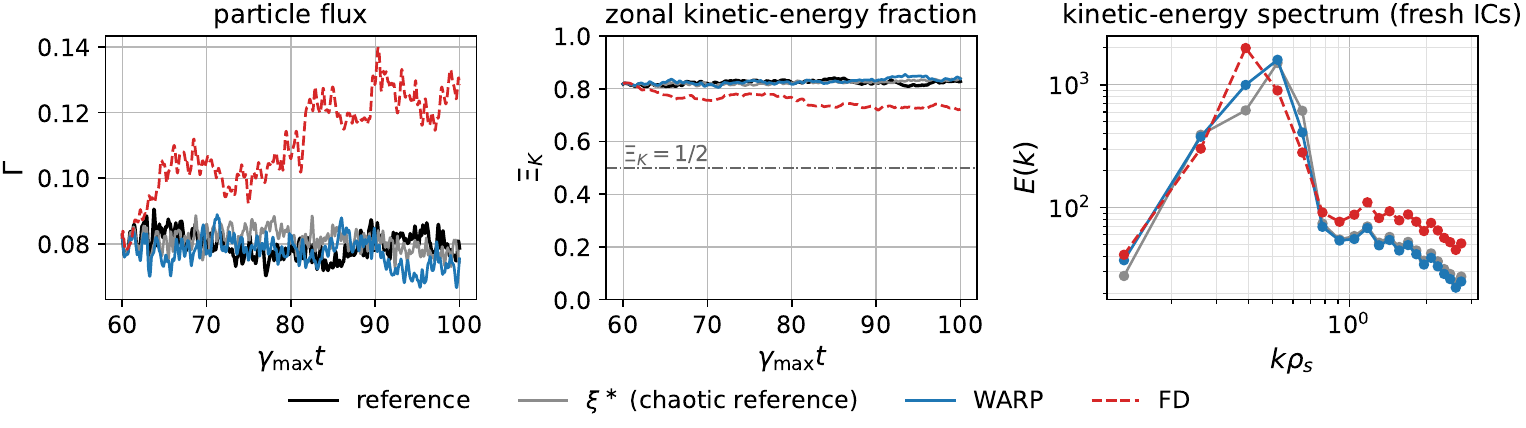}
\caption{mHW at $C=1$ and $\eta=5\%$: particle flux and zonal kinetic-energy fraction along the deployment of Fig.~\ref{fig:mhw_errtime}, and the kinetic-energy spectrum averaged over the fresh initial conditions of Fig.~\ref{fig:mhw_stat_heldout}}
\label{fig:mhw_deploy_diag}
\end{figure}

Figure~\ref{fig:mhw_deploy_diag} shows the same deployments on the diagnostics used to validate the data: FD runs a persistently high flux, lets the zonal fraction drift downward, and piles kinetic energy at the top of the band, whereas the WARP traces and spectrum are indistinguishable from the true model.

\subsection{Decaying MHD and the Orszag--Tang vortex (2D)}
\label{sec:ot2d}
The final benchmark is a deterministic MHD transient. In the Orszag--Tang vortex~\citep{OrszagTang1979}, smooth velocity and magnetic fields steepen into thin current sheets, and solver quality is conventionally assessed through peak current and dissipation~\citep{Rosenberg2007}. The five-term vorticity--flux system and its coefficients are listed in Table~\ref{tab:settings}. The discovery window $t\in[0.15,1.5]$ covers the current-sheet event.

Here $u=(\chi_y,-\chi_x)$, $\omega=-\Delta\chi$, $B=(A_y,-A_x)$, and $j=-\Delta A$ are velocity, vorticity, magnetic field, and current density. The diagnostics $E_u$ and $E_B$ are kinetic and magnetic energy, $E_{\rm tot}=E_u+E_B$ is total energy, and $\varepsilon_{\rm tot}$ is the sum of viscous and resistive energy dissipation. For the evaluation interval $\mathcal T$ of each deployment, define the dissipation-peak time and maximum current density by
\begin{equation}
t^\ast:=\underset{t\in\mathcal T}{\operatorname{arg\,max}}\,\varepsilon_{\rm tot}(t),
\qquad
J_{\max}(t):=\|j(\cdot,t)\|_\infty,
\qquad J^\ast:=\max_{t\in\mathcal T}J_{\max}(t).
\label{eq:ot_events}
\end{equation}
The corresponding relative event errors are
\begin{equation}
e_{t^\ast}:=\frac{|t^\ast-t^\ast_{\rm ref}|}{t^\ast_{\rm ref}},
\qquad
e_{J^\ast}:=\frac{|J^\ast-J^\ast_{\rm ref}|}{J^\ast_{\rm ref}}.
\label{eq:ot_event_errors}
\end{equation}
\begin{figure}
\centering
\includegraphics[width=\linewidth]{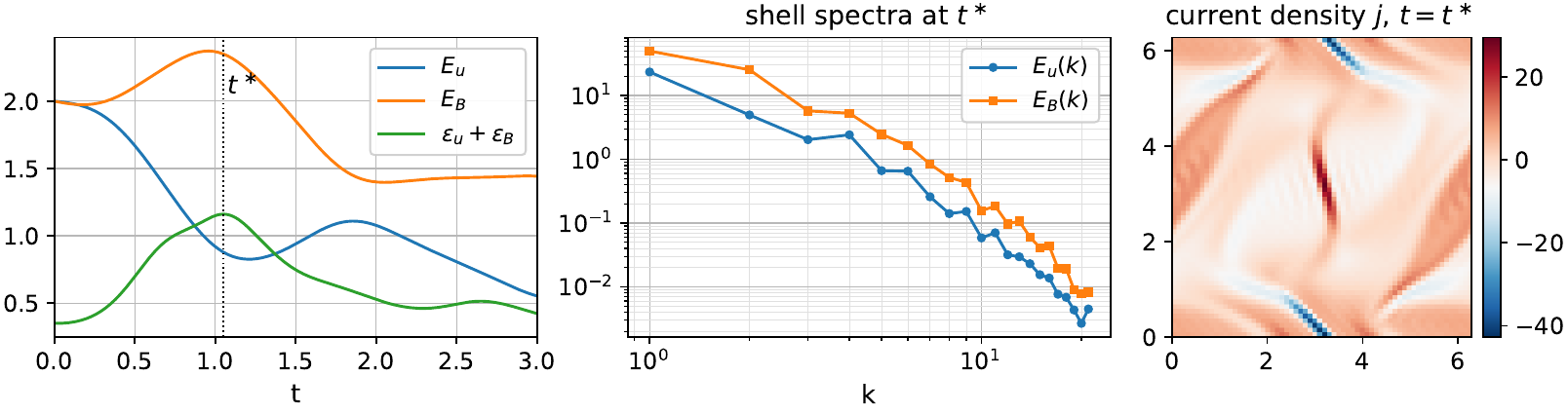}
\caption{Orszag--Tang data-generation diagnostics on the $64^2$ grid with $M=21$; $t^\ast$ is the dissipation-peak time of Eq.~\eqref{eq:ot_events}}
\label{fig:ot_sanity}
\end{figure}
The subscript $\mathrm{ref}$ denotes the true-coefficient model integrated from the same fresh initial condition and over the same interval as the learned model. The current maximum need not occur at the dissipation-peak time. These event diagnostics are used only for Orszag--Tang.

\begin{figure}
\centering
\includegraphics[width=0.45\linewidth,keepaspectratio]{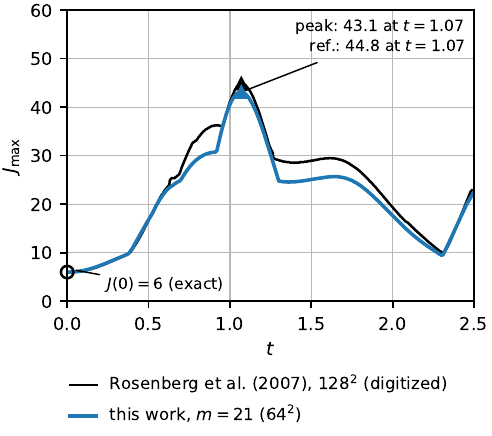}
\caption{Solver validation of $J_{\max}(t)$ for the $M=21$ generator in blue and the $128^2$ reference of \citet{Rosenberg2007} in black}
\label{fig:ot_m1}
\end{figure}

Figure~\ref{fig:ot_sanity} shows the data-generation diagnostics. The transient exchanges kinetic for magnetic energy while the total decays, the joint dissipation peaks at $t^\ast=1.05$, and the current density exhibits a central sheet. Figure~\ref{fig:ot_m1} compares $J_{\max}(t)$ with the $128^2$ pseudo-spectral reference of \citet{Rosenberg2007}. The first-peak time agrees and the $M=21$ peak amplitude is $4\%$ low.

\begin{figure}
    \centering
    \begin{subfigure}[b]{0.4\linewidth}
        \centering
        \includegraphics[width=\linewidth,keepaspectratio]{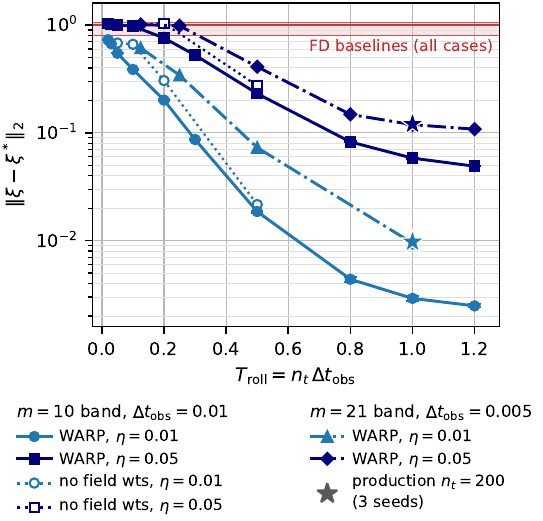}
        \caption{}
        \label{fig:ot_troll}
    \end{subfigure}
    \hfill
    \begin{subfigure}[b]{0.55\linewidth}
        \centering
        \includegraphics[width=\linewidth,keepaspectratio]{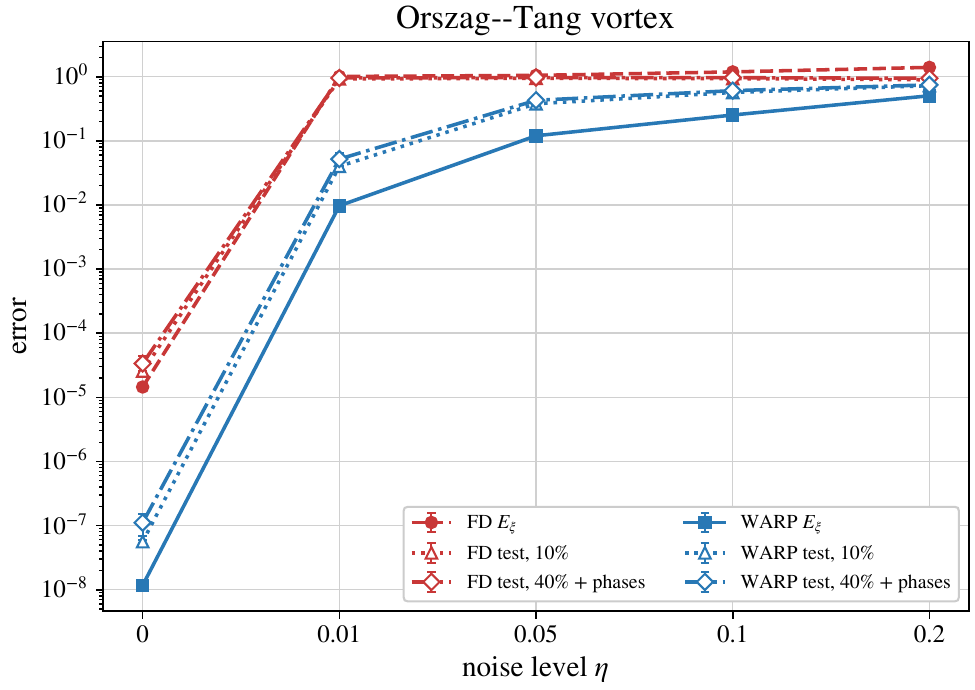}
        \caption{}
        \label{fig:ot_noise}
    \end{subfigure}
    \caption{Orszag--Tang robustness studies.
    (a)~Horizon sweep: $E_\xi$ versus rollout duration $T_{\rm roll}=H\Delta t_{\rm s}$. Open markers omit the per-field weights of Sec.~\ref{sec:warp_training}; stars mark the production setting ($H=200$), averaged over three noise seeds, and all other points use seed~42. 
    (b)~Coefficient error $E_\xi$ and testing trajectory error versus noise level}
    \label{fig:ot_noise_troll}
\end{figure}

Production training uses a 200-step forward rollout. Because vorticity has roughly twice the rms amplitude of the magnetic flux function, the two state blocks receive inverse-rms weights. Figure~\ref{fig:ot_troll} shows the retained sensitivity tests for rollout duration and field weighting.

Figure~\ref{fig:ot_noise} combines coefficient recovery and testing trajectory errors. Under noise, FD first loses the Lorentz coefficient because the current-density column emphasizes noisy high wavenumbers, followed by induction advection. WARP preserves the leading advection and Lorentz coefficients and degrades gradually. Its remaining high-noise error concentrates in the dissipative coefficients. Its deployed model places the dissipation peak and current maximum within a few percent at low noise.

\begin{figure}
\centering
\includegraphics[width=0.9\linewidth,keepaspectratio]{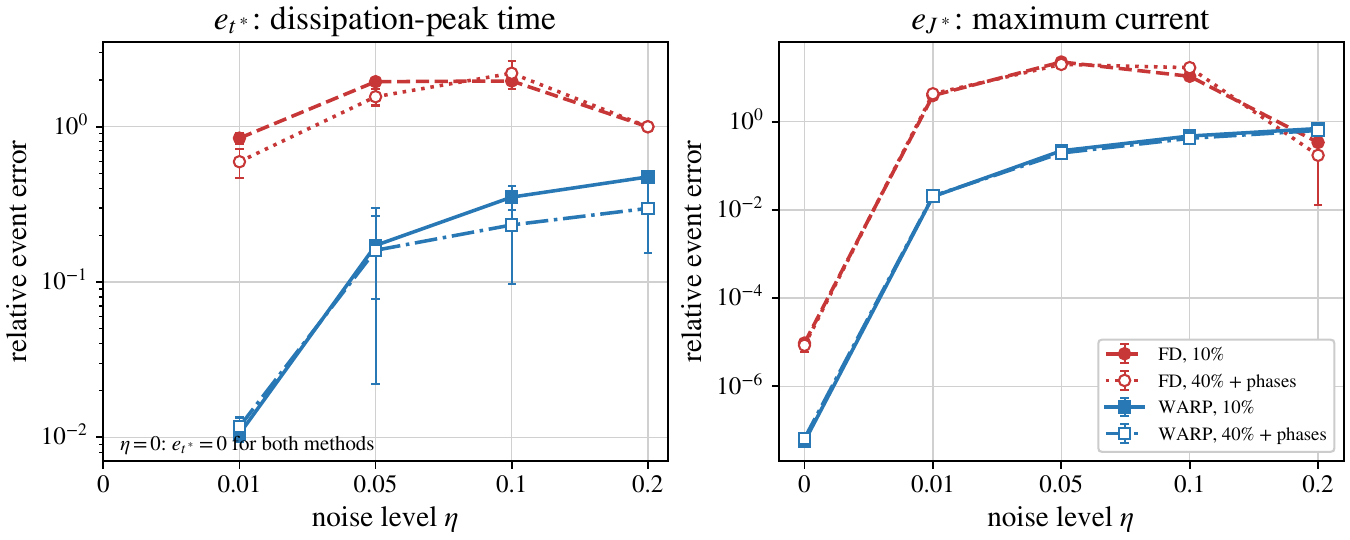}
\caption{Orszag--Tang event errors of Eq.~\eqref{eq:ot_event_errors} from five fresh initial conditions, using the seed-42 fit. Error bars show one standard deviation over the initial conditions}
\label{fig:ot_heldout}
\end{figure}

\begin{figure}
\centering
\includegraphics[width=0.95\linewidth,keepaspectratio]{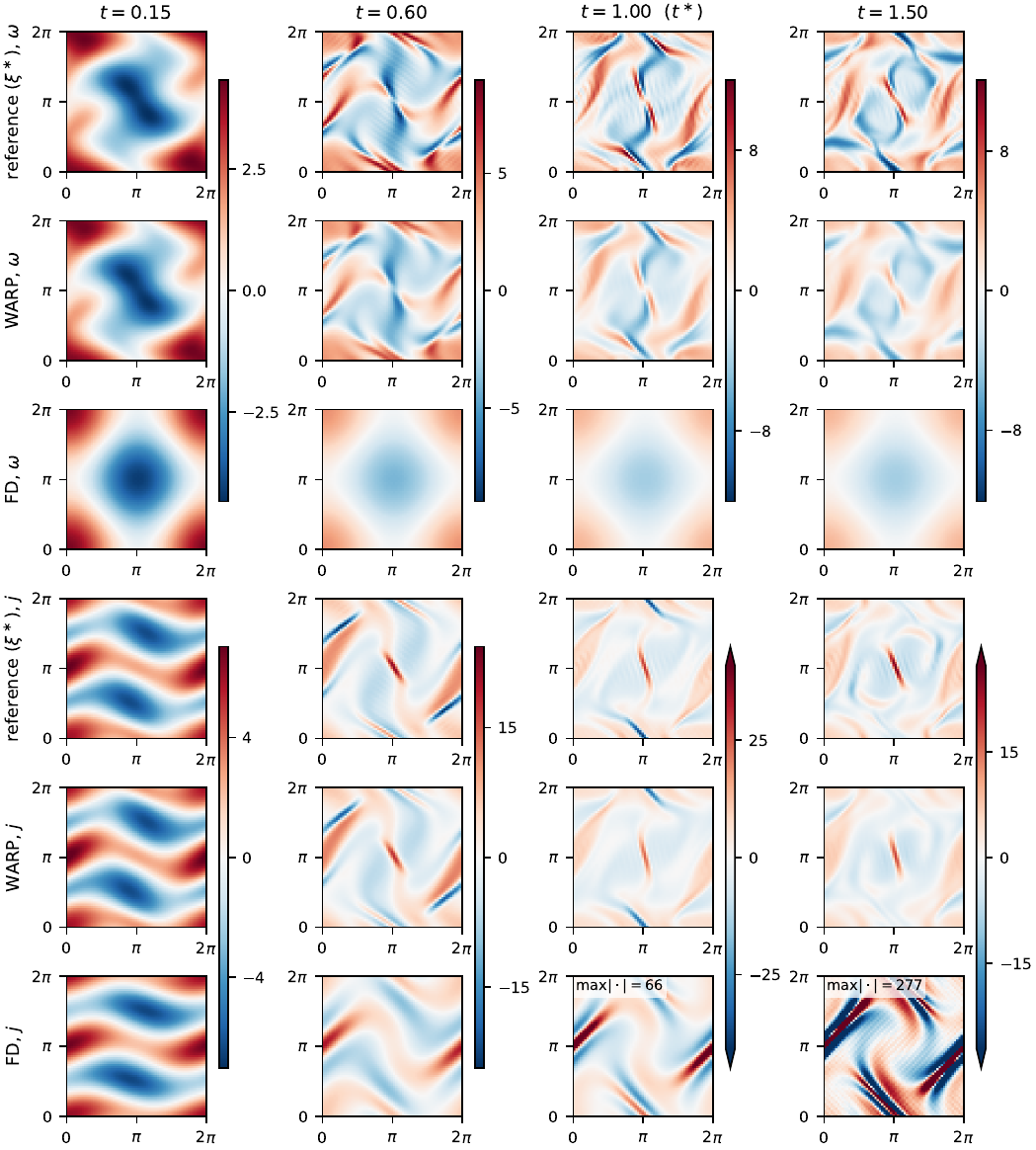}
\caption{Orszag--Tang deployment from a held-out initial condition on the production band with a $64^2$ grid and $M = 21$, for coefficients learned at $\eta = 5\%$. Rows within each block show the true model, the WARP deployment, and the FD deployment; the blocks are vorticity $\omega$ and current density $j$. All three are integrated from the same new initial condition. Columns span the discovery window and include the current-sheet event of this initial condition, labeled $t^\ast$ in the panel titles. Color scales follow the reference at each time; panels that exceed the range are annotated with their peak}
\label{fig:ot_fields_heldout}
\end{figure}

We also deploy the learned coefficients from five fresh initial conditions, using $\pm10\%$ amplitude perturbations and a $\pm40\%$ fresh-phase stress test (Fig.~\ref{fig:ot_heldout}). WARP's event errors grow from a few percent at $\eta=1\%$ to order one at $\eta=20\%$, where the event is no longer resolved, whereas FD shifts the dissipation peak by an amount comparable to $t^\ast$ itself from $1\%$ noise onward and overestimates the peak current by a factor of five to twenty-four for $\eta\le0.1$. The small FD current-error values at $\eta=0.2$ occur after the model has lost the current-sheet event and are not evidence of an accurate trajectory (see the trajectory errors in Fig.~\ref{fig:ot_noise}). Figure~\ref{fig:ot_fields_heldout} shows both fields across the window for a held-out initial condition. WARP reproduces the vorticity and current-density structure of the reference at every time, including the current sheet at $t^\ast$, with the same peak amplitude to within the band truncation. FD reproduces the vorticity but its current density leaves the reference color range at and after the event, with the overestimated peak annotated in the panels. Matching the $128^2$ reference's sup-norm amplitude exactly would require a wider retained band.

\FloatBarrier
\subsection{Robustness checks}
\label{sec:robustness}

\paragraph{Sensitivity to discretization choices.}\label{sec:consistency_experiments}
The Galerkin formulation leaves only two discretization choices, the number of retained modes and the training horizon $H$.
Sweeping the ZK basis size and the Burgers horizon (Appendix~\ref{app:supp_results}) shows that coefficient recovery is statistically flat in the number of retained modes, with no truncation floor arising because discovery and data generation share the discretization, and flat in $H$ beyond $H\approx25$, so neither behaves as a delicate tuning parameter and the default $H=50$ lies on the plateau.

\section{Limitations and Future Work}\label{sec:limitations}

The FD method deliberately shares WARP's Galerkin representation, preprocessing, and library, so the present comparison isolates the effect of multistep rollout; Appendix~\ref{app:biglib} relaxes the library to an overcomplete one and reaches the same conclusion. A strong-form control on the same trajectories would be required to quantify the separate contribution of weak spatial projection. Comparisons with other weak-form implementations would mix training criteria with different test functions and discretizations.

The experiments assume periodic domains and candidate libraries containing the correct operators. Larger libraries require sparsity promotion or stability selection~\citep{Stability,BlumensathDavies2009}, and nonperiodic domains require an appropriate Galerkin basis. An obvious extension for addressing more general PDEs and boundary conditions is to adopt finite-element-style local basis functions. Experimental noise, missing fields, and shocks also remain untested; discontinuities can reduce weak-quadrature accuracy~\citep{Gurevich2019}, while laboratory and turbulent-flow studies show that weak formulations remain promising beyond the synthetic Gaussian-noise setting~\citep{Reinbold2021,Gurevich2024SPIDER}.

The turbulent studies have additional limits. The independent DNS projection is unclosed, but the library does not yet include closure terms. The truncated, resolution-matched system captures the qualitative mHW transition with a quantitative offset. The Orszag--Tang current-sheet amplitude is limited by the $M=21$ band. 

\section{Conclusion}
\label{sec:conclusion}

WARP identifies interpretable PDE coefficients by differentiating through anchored multistep trajectories of the same weak Galerkin model used for deployment. The learned coefficients define an autonomous spectral solver that supports direct forward integration.

Across eight fluid and plasma systems, WARP improves coefficient recovery, rollout accuracy, and prediction of statistical measures over derivative fitting on the identical weak formulation. The advantage persists for coupled turbulence and a current-sheet-forming MHD transient, and in coefficient error for unclosed $512^2$ DNS projections. These results also hold under strong noise, although future work must validate this conclusion when the dynamics are not known and/or when using a large library of candidate differential-operator terms.

The approach therefore separates two complementary sources of robustness. Spatial weak projection limits derivative amplification of observation noise, as established in prior weak-form work, and multistep rollout, whose contribution the present comparison isolates, constrains and strongly improves the accuracy of the integrated dynamics. Extending this combination to automated term selection, large operator libraries, nonperiodic and local basis functions, and a profiled FFT implementation for larger modal systems is the next step.

\section*{Acknowledgments}
This work was supported by the National Science Foundation (NSF) under Award No.\ 2512173. We thank Elizabeth Paul and Weichao Tu for useful discussions.

\section*{Declaration of generative AI and AI-assisted technologies in the writing process}
Large language models developed by Anthropic PBC and OpenAI PBC were used throughout this work as coding assistants, for implementing and refactoring parts of the WARP code base. They were also used to proofread the manuscript and to check derivations. All code and text were reviewed and edited by the authors, who take full responsibility for the content of this article.
% \newpage
\begin{appendices}
\makeatletter
\def\@fb@secFB{}
\makeatother
\section{Additional Hasegawa--Wakatani Diagnostics}
\label{app:hw_additional}
The main text shows the $c_1=1$ identification and validation results. This appendix retains the other adiabaticities and the spatial snapshots of the $c_1=1$ deployment.

\begin{figure}
\centering
\includegraphics[width=\linewidth]{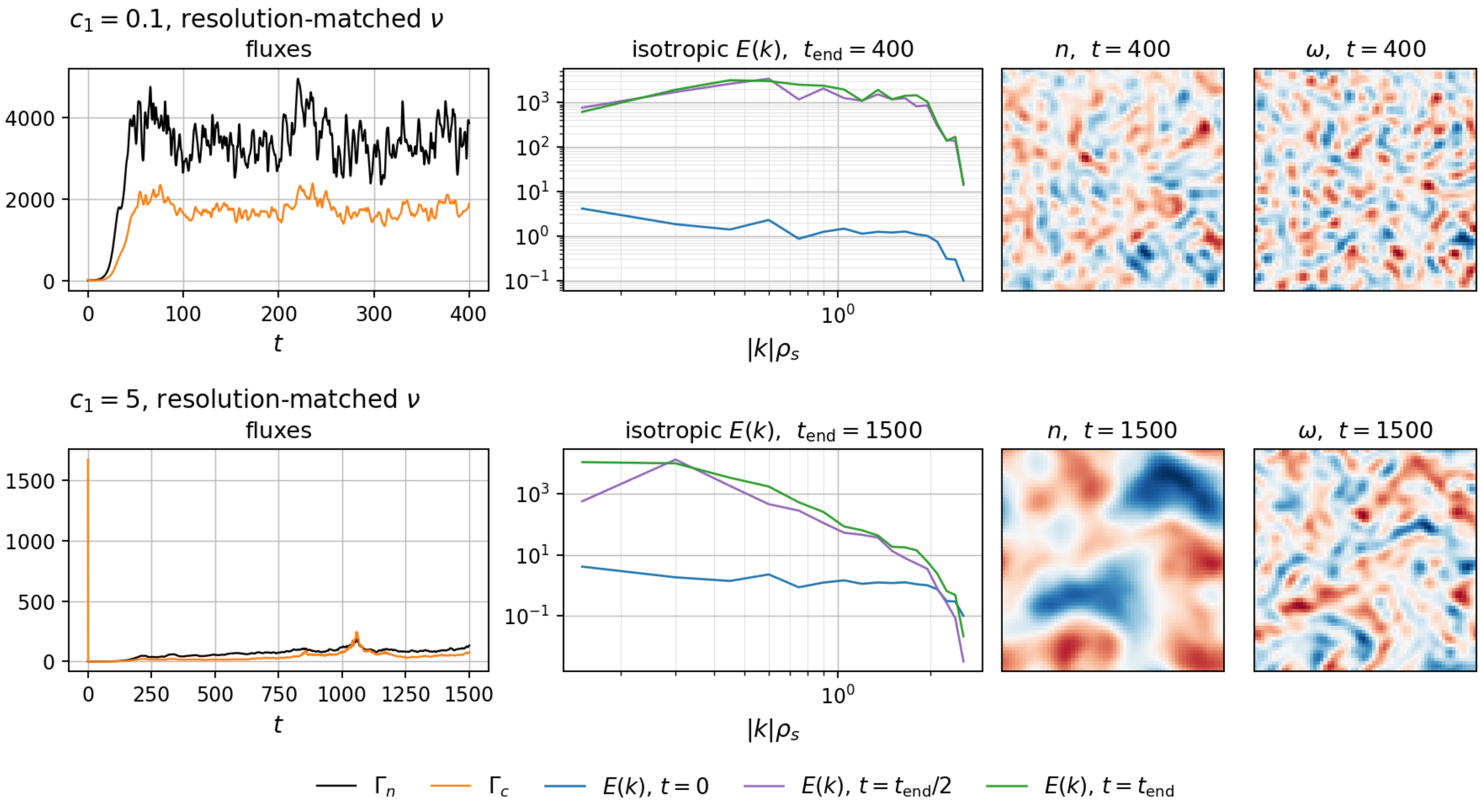}
\caption{Additional band-closed HW data-generation diagnostics at $c_1=0.1$ (top) and $c_1=5$ (bottom), using resolution-matched hyperdiffusion. Columns and colors follow Fig.~\ref{fig:hw_turb_sanity}: $\Gamma_n$ is particle flux, $\Gamma_c$ resistive energy dissipation, $E(k)$ shell energy, $n$ density fluctuation, and $\omega$ vorticity. Both flux traces are spatially integrated}
\label{fig:hw_turb_sanity_other}
\end{figure}

\begin{figure}
\centering
\begin{subfigure}[t]{0.49\linewidth}
\centering
\includegraphics[width=\linewidth]{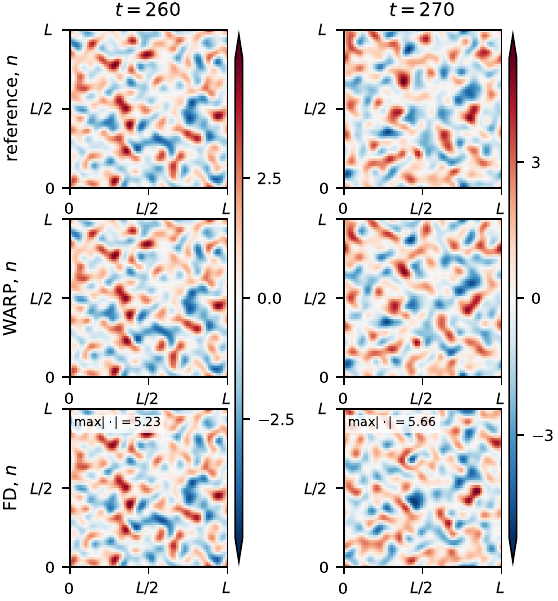}
\caption{density $n$}
\end{subfigure}\hfill
\begin{subfigure}[t]{0.49\linewidth}
\centering
\includegraphics[width=\linewidth]{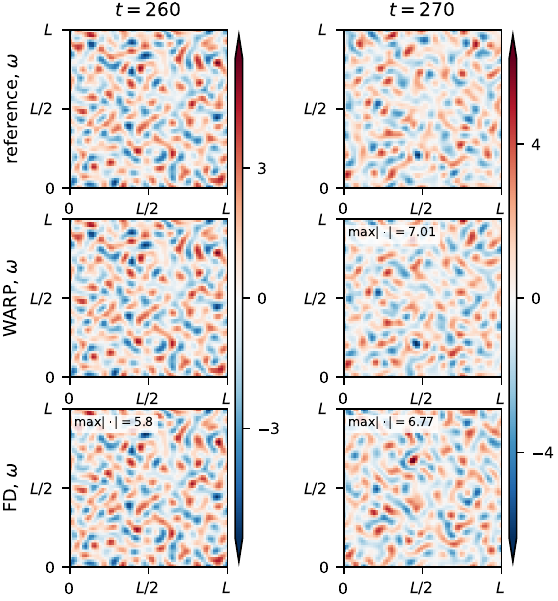}
\caption{vorticity $\omega$}
\end{subfigure}
\caption{Saturated HW turbulence at $c_1=1$ and $\eta=5\%$. Rows show the clean reference and the WARP and FD deployments launched from the noisy snapshot at $t=250$; columns are $10$ and $20$ time units ($20$ and $40$ training horizons) after launch. Color scales are set by the reference at each time and shared down a column; panels that leave the range are annotated with their peak}
\label{fig:hw_turb_rollout}
\end{figure}

\begin{figure}
\centering
\includegraphics[width=\linewidth]{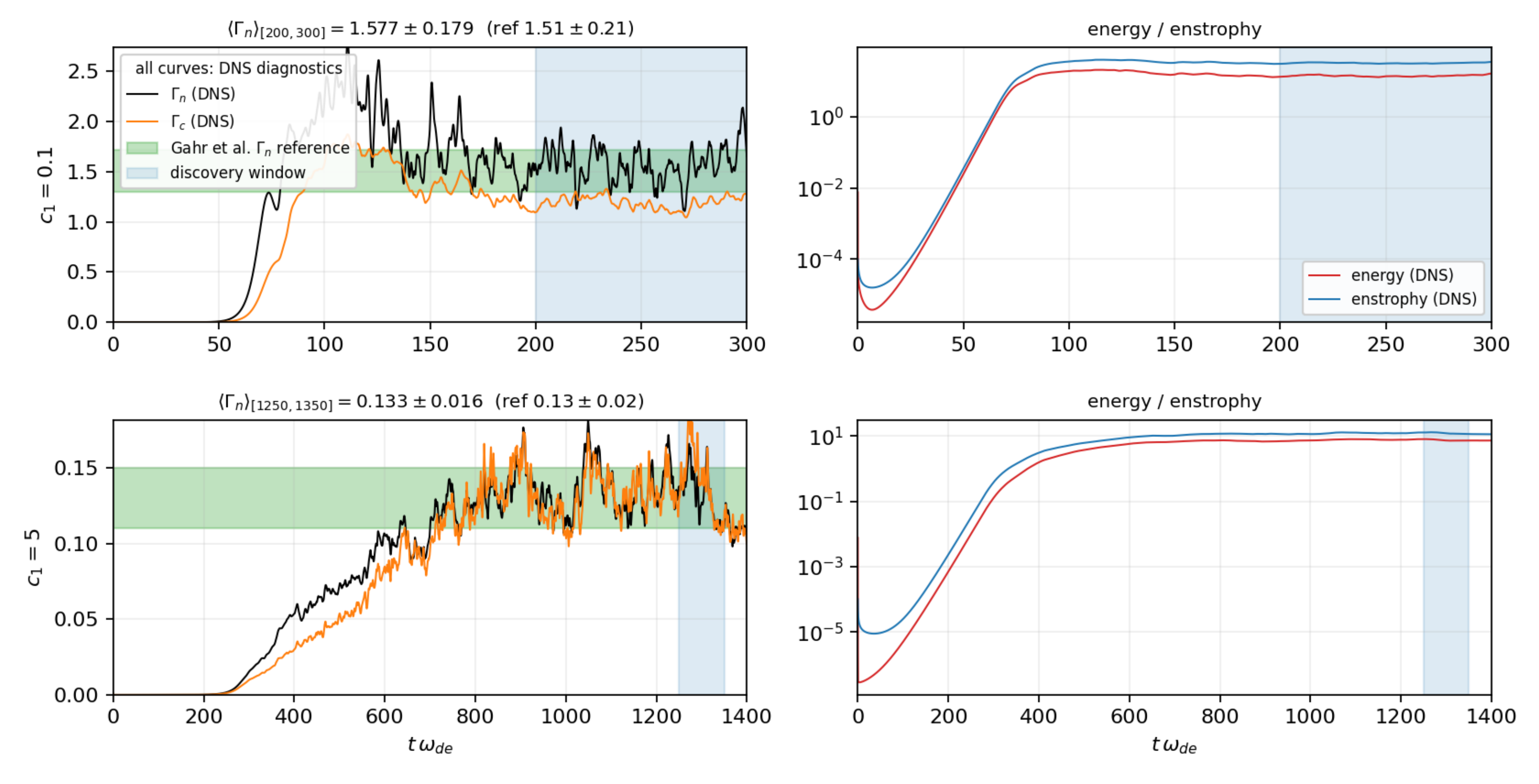}
\caption{As Fig.~\ref{fig:dns512_traces}, for $c_1=0.1$ (top) and $c_1=5$ (bottom)}
\label{fig:dns512_traces_other}
\end{figure}

\begin{figure}
\centering
\includegraphics[width=0.65\linewidth]{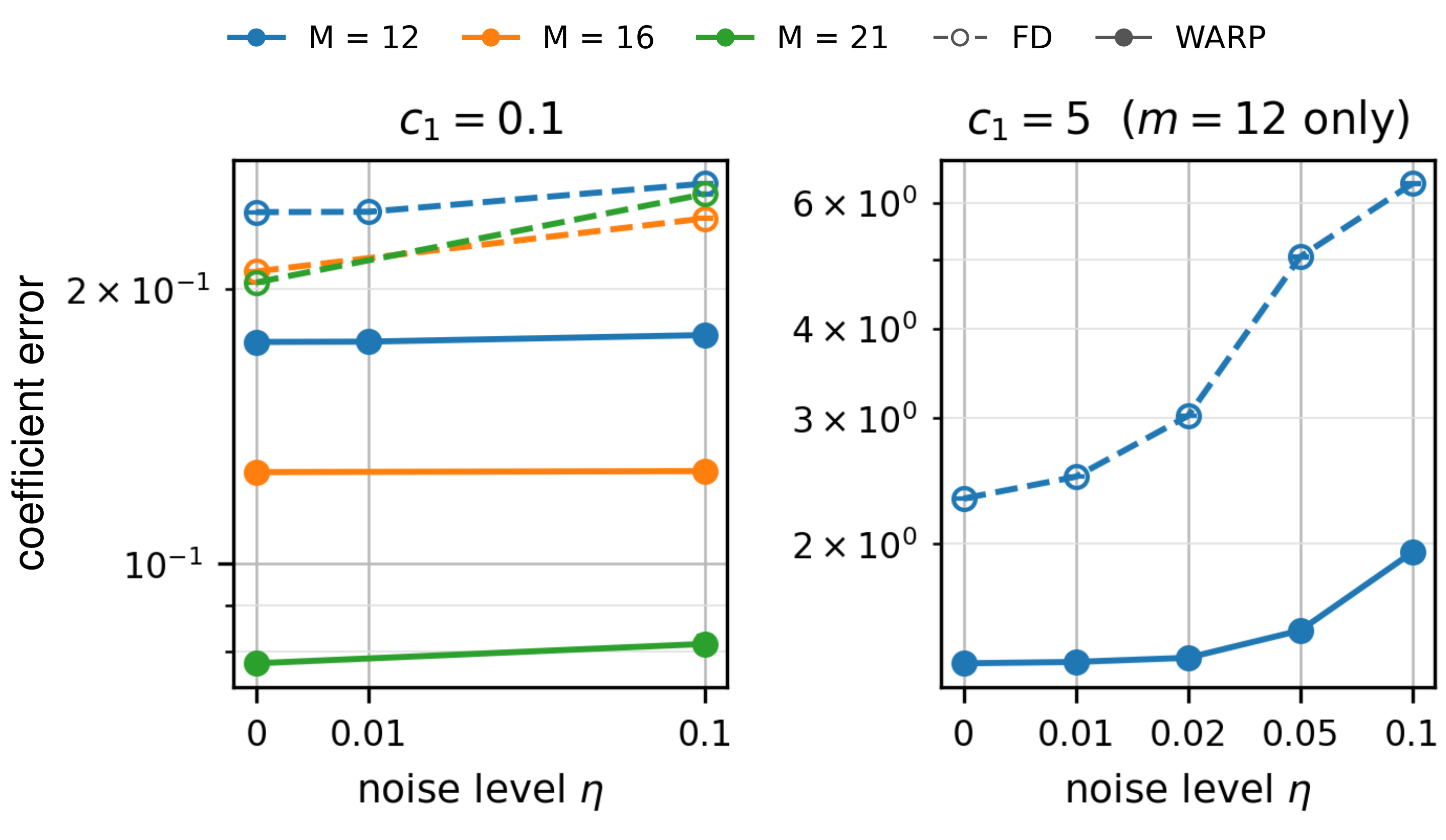}
\caption{As Fig.~\ref{fig:dns512_closure}, identification from projected independent DNS at $c_1=0.1$ (left) and $5$ (right)}
\label{fig:dns512_closure_other}
\end{figure}

\FloatBarrier
\section{Mechanisms of the Rollout Advantage}
\label{sec:discussion}

\subsection{Why multistep rollout training helps}
\label{sec:mechanism}

Finite temporal offsets carry more signal relative to observation noise than an instantaneous finite-difference derivative. For small $r\Delta t_{\rm s}$, the signal in $q_{\mathrm{data}}(t_{j+r})-q_{\mathrm{data}}(t_j)$ grows approximately linearly with $r$, while the difference of independent snapshot noise remains offset-independent. WARP also evaluates the weak library on its predicted state at every right-hand-side call. A coefficient vector that matches local increments but drifts under integration is therefore penalized directly, and all offsets are coupled through a single model trajectory. This integrated test provides benefits beyond denoising. It rejects many coefficient vectors that lead to unstable or rapidly diverging forward solutions and therefore improves the chance that a learned model can be deployed from a new initial condition~\citep{LoiseauBrunton2018,KaptanogluTrapping2021,GoyalBenner2022,LeeTraskStinis2022,BannaUnrolled2027}.

\subsection{Coefficient recovery and noise sensitivity}
\label{sec:coeff_noise}

Across the 1D benchmarks, WARP's advantage is modest for Burgers and reaches 1--2 orders of magnitude for KdV, whose dispersive column weights the highest retained modes as $k^3$ (Sec.~\ref{sec:kdv_experiments}). This is an empirical trend and not a scaling law because the problems also differ in resolution and time scale.

FD degrades gradually on Burgers, ZK, Navier--Stokes, and HW, but becomes dominated by noise on KdV, Kolmogorov flow, and the Orszag--Tang vortex. Once FD is saturated, larger noise primarily increases WARP's error and therefore reduces the ratio between methods even though WARP remains substantially more accurate. In Kolmogorov flow and mHW the central-difference estimator carries a separate sampling-rate bias (Appendix~\ref{app:sampling_checks}).

\section{Sampling-Rate and Resolution Checks}
\label{app:sampling_checks}

\paragraph{Kolmogorov-flow resolution.}
The identification run begins from the laminar streamfunction plus a fixed low-mode perturbation and discards 500 time units of spin-up. Let $E_{\rm lam}$ and $D_{\rm lam}$ denote the energy and dissipation of the laminar Kolmogorov solution. On the $64^2$, $M=21$ band, energy saturates near $0.44E_{\rm lam}$, dissipation and injection agree near $0.09D_{\rm lam}$, and the spectrum decays monotonically beyond the forcing wavenumber $k_f=4$ without band-edge pile-up. A $128^2$, $M=42$ companion changes mean energy by $0.34\%$ and leaves mean dissipation within block-sampling error (Fig.~\ref{fig:kolm_sanity}).

\paragraph{mHW validation.}
Each adiabaticity uses a box containing ten fastest-growing wavelengths and resolution-matched dissipation in both fields. A small-amplitude unstable eigenmode reproduces the dispersion-relation growth rate within $2\%$. 

\paragraph{Sampling-rate bias.}
FD has $O(\Delta t_{\rm s}^2)$ truncation error on each mode, so a sampling interval that does not resolve a dynamically important rate biases the fit even without observation noise. In mHW, the resistive coupling relaxes the density--potential difference at rate $C(1+|k|^{-2})$ for a nonzonal Fourier mode with wavevector $k$ (ignoring drive and diffusion). This rate is largest at the smallest nonzero wavenumbers, where the state can carry substantial amplitude. If the observation interval does not resolve that relaxation, a central difference biases an influential part of the regression. For Kolmogorov flow the viscous decay rate instead grows as $\nu|k|^2$; the fastest modes lie at high wavenumber, where the measured spectrum is weak, so their under-resolution has a smaller effect on the fit. WARP integrates the candidate model adaptively between observations, which decouples integration accuracy from the sampling rate.

\section{Hyperparameter Sensitivity}
\label{app:supp_results}

\begin{figure}[!htbp]
\centering
\includegraphics[width=0.7\linewidth]{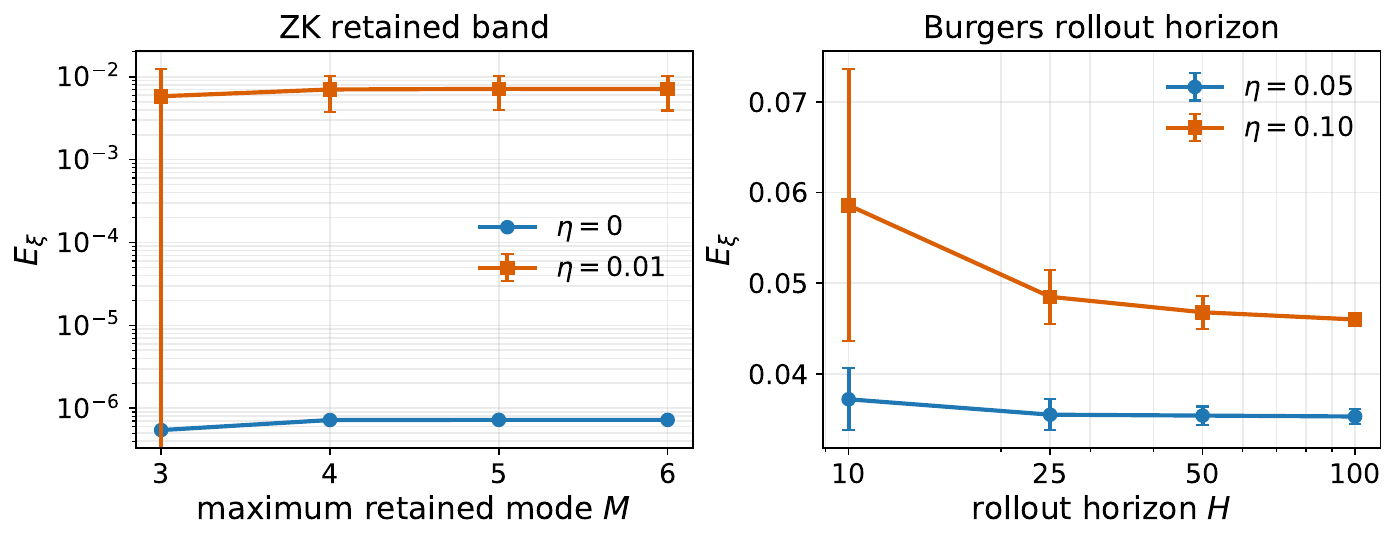}
\caption{WARP coefficient error versus retained band $M$ (ZK) and rollout horizon $H$ (Burgers). Error bars show one standard deviation over three noise seeds}
\label{fig:hyperparameter_sensitivity}
\end{figure}

Figure~\ref{fig:hyperparameter_sensitivity} reports the retained sensitivity runs. The ZK coefficient error is statistically flat across the band sweep, and the Burgers error changes little once $H\ge25$.

\section{Long-Horizon Deployment}
\label{app:longroll}

\begin{figure}[!htbp]
\centering
\begin{subfigure}{\linewidth}
  \centering
  \includegraphics[width=\linewidth]{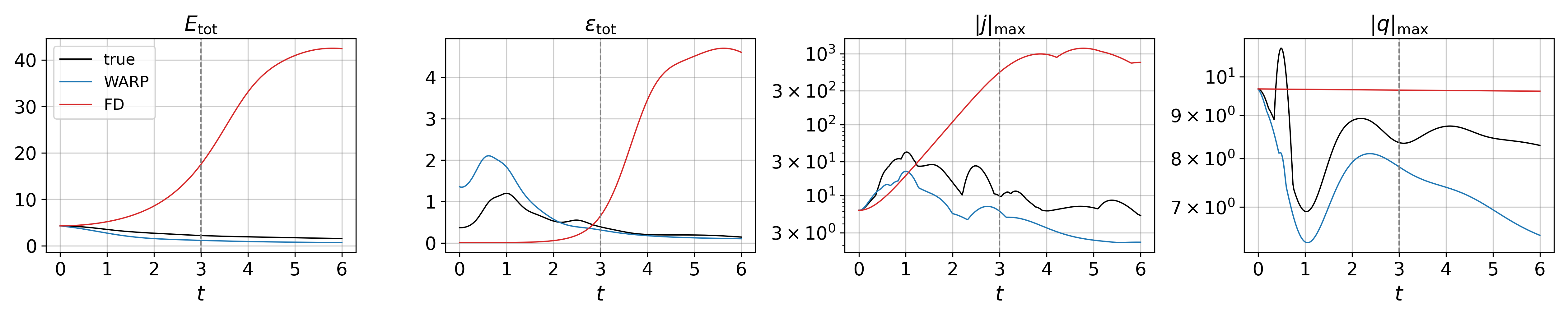}
  \caption{Orszag--Tang}
  \label{fig:longroll-ot}
\end{subfigure}

\begin{subfigure}{\linewidth}
  \centering
  \includegraphics[width=\linewidth]{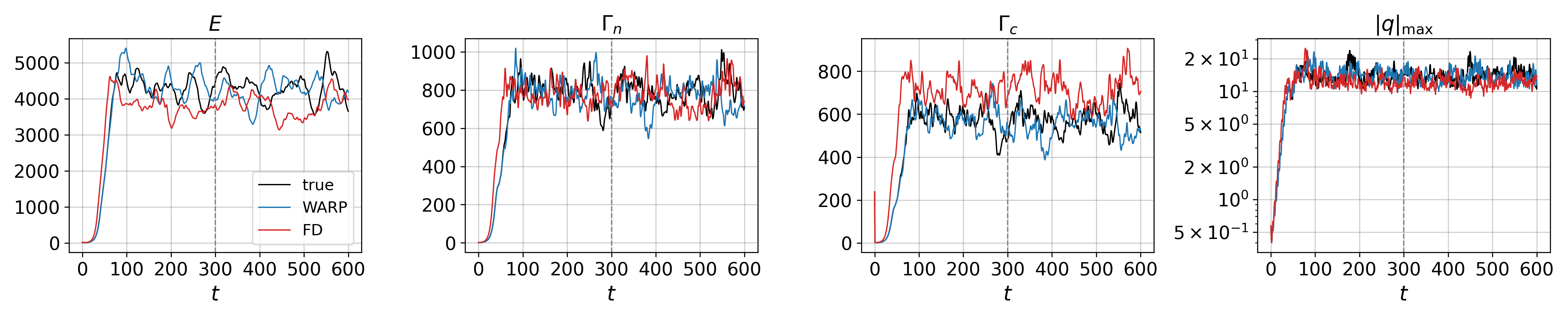}
  \caption{Hasegawa--Wakatani}
  \label{fig:longroll-hw}
\end{subfigure}

\begin{subfigure}{\linewidth}
  \centering
  \includegraphics[width=\linewidth]{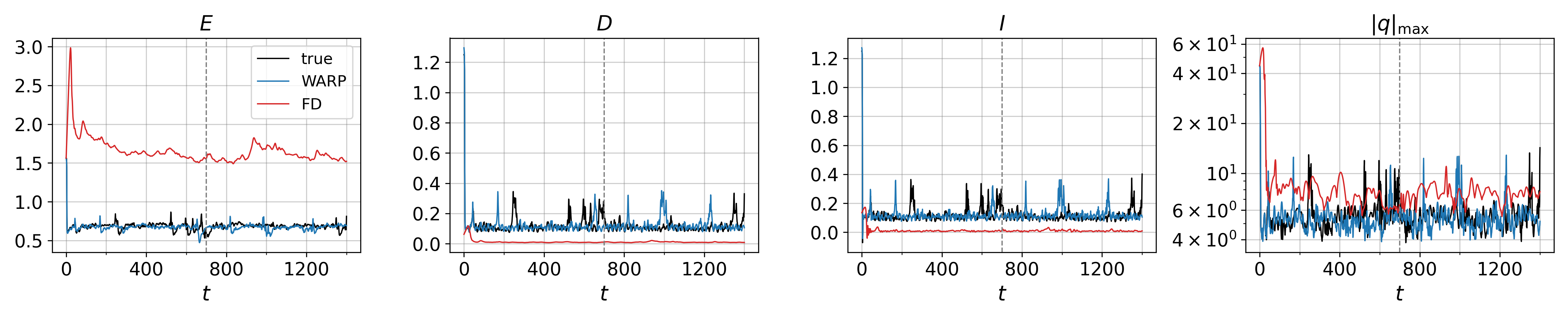}
  \caption{Kolmogorov flow}
  \label{fig:longroll-kolm}
\end{subfigure}
\caption{Long-horizon deployment of the $\eta=10\%$ models from a new initial condition, for twice the original testing interval (dotted line). HW uses $c_1=1$; $|q|_{\max}$ is the largest absolute modal coefficient}
\label{fig:longroll}
\end{figure}

To probe stability beyond the deployment windows of Sec.~\ref{sec:experiments}, we integrate the WARP and FD models trained at $\eta=10\%$ from an initial condition not used during training, over twice the original held-out horizon and with the same adaptive integrator. The reference trajectory is obtained by integrating the true model $\xi^\ast$ from the same initial condition (Fig.~\ref{fig:longroll}).

WARP stays stable and follows the reference in all three systems. In Orszag--Tang, the energy decays monotonically as in the reference but too quickly, since viscosity and resistivity are overestimated while the advection and Lorentz coefficients are accurate. In HW the saturated fluxes match the reference, and in Kolmogorov flow the energy budget is reproduced, intermittent dissipation bursts included.

FD remains numerically bounded in every case, but its failure mode depends on each system's intrinsic dissipation. Orszag--Tang has none beyond the learned viscosity and resistivity, and FD suppresses both along with essentially all Lorentz coupling. The magnetic field is then advected passively, and the total energy, which should decay, instead grows by an order of magnitude. In HW, hyperdiffusion and the adiabatic coupling keep FD bounded, but its resistive dissipation is biased high. In Kolmogorov flow, the fitted coefficients shrink by roughly the same factor, consistent with errors-in-variables bias~\citep{BortzWENDy2023,MessengerBortzConsistency2025}, so the FD dynamics run about an order of magnitude too slowly and never complete the laminar-to-turbulent transition within the window. Over long horizons, FD thus either develops spurious energy growth or settles on the wrong attractor, while WARP stays on the correct one. Strong-form identification can also produce unstable rollouts~\citep{KaptanogluBenchmark2023}, but no strong-form control is performed here, so these experiments do not establish that weak projection alone prevents blow-up. 

\section{Identification with an Overcomplete Library}
\label{app:biglib}

\begin{figure}[!htbp]
\centering
\includegraphics[width=0.85\linewidth]{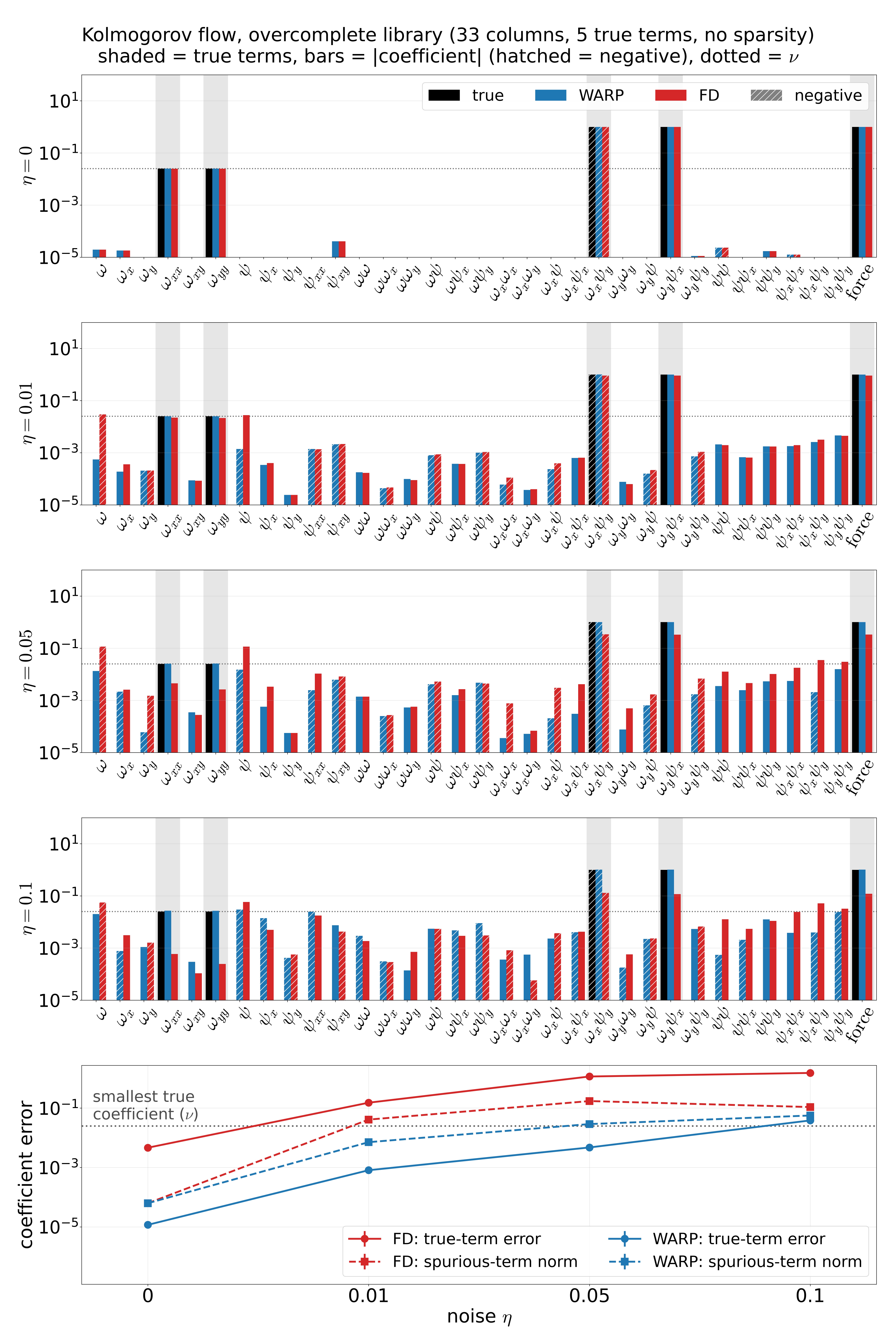}
\caption{Kolmogorov flow with the overcomplete library: recovered coefficient magnitudes at each noise level (top four rows), and true-term error and spurious-term norm versus noise (bottom)}
\label{fig:kolm_biglib}
\end{figure}

The libraries of Sec.~\ref{sec:experiments} contain \emph{exactly} the correct terms of the governing equations. To check that the rollout advantage survives without this prior knowledge, we repeat the Kolmogorov benchmark with a library of all terms up to quadratic in the fields and up to second order in spatial derivatives: the linear terms in $\omega$ and $\psi$ and their first and second derivatives, all pairwise products of $\omega$, $\psi$, and their first derivatives, and the forcing, 33 columns in total. Every column is projected onto zero mean, since the periodic vorticity equation only admits right-hand sides with zero mean; the true terms already satisfy this. Five columns belong to the true model. The advection appears as the antisymmetric pair $\omega_x\psi_y$, $\omega_y\psi_x$, the Laplacian as the two columns $\partial_{xx}\omega$ and $\partial_{yy}\omega$, and the forcing once. Figure~\ref{fig:kolm_biglib} shows the recovered coefficients together with the error on the true terms and the norm of the spurious ones.

On clean data, both methods recover the full coefficient vector. With noise, the two methods separate in the same way as with the minimal library. FD attenuates the true coefficients by the same factors as before, so enlarging this particular library does not remove the attenuation associated with noisy regressors~\citep{BortzWENDy2023,MessengerBortzConsistency2025}. FD also puts weight on the spurious columns at every noise level, mostly on a pair of linear terms in $\omega$ and $\psi$ with opposite signs, which amounts to a linear damping at high wavenumber in place of the attenuated viscosity. Above one percent noise this spurious weight is larger than FD's viscosity coefficient, so classical sparsity-promoting methods (e.g. with thresholding) would produce the wrong terms.

WARP recovers the true coefficients to within a few percent at all noise levels and keeps the spurious coefficients an order of magnitude smaller than FD's. At the lowest noise its spurious coefficients agree term by term with FD's, so they are the bias that observation noise leaves in the quadratic columns. They grow with noise, and at ten percent the largest of them are as large as the recovered viscosity. 

\begin{figure}[!htbp]
\centering
\includegraphics[width=\linewidth]{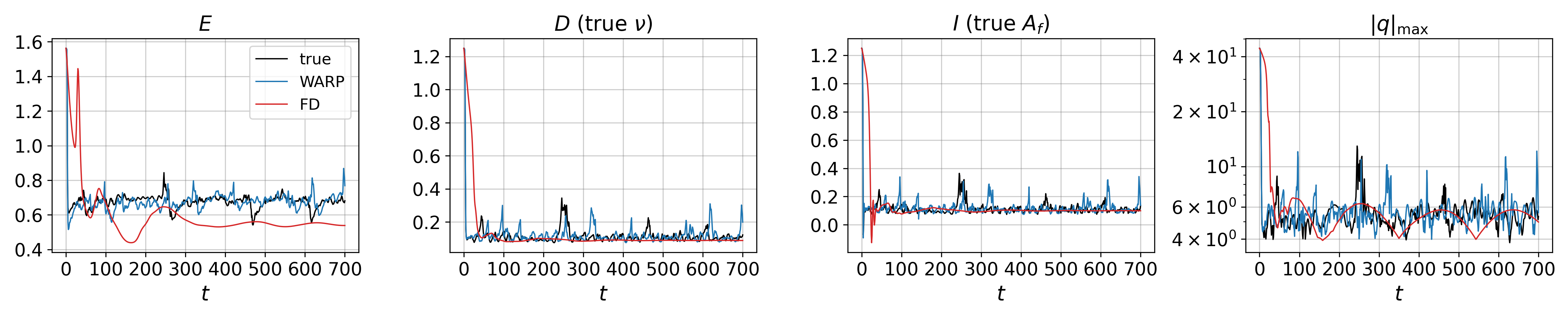}
\caption{Deployment of the overcomplete-library models trained at $\eta=10\%$ from a new initial condition. $D$ and $I$ are evaluated with the true $\nu$ and $A_f$ for all models}
\label{fig:kolm_biglib_roll}
\end{figure}

Deployed from a new initial condition for the original held-out horizon, both models trained at $\eta=10\%$ remain bounded (Fig.~\ref{fig:kolm_biglib_roll}). WARP reproduces the energy budget of the true model, including the intermittent dissipation bursts, and its spurious coefficients do not destabilize the rollout. FD, by contrast, gets the dynamics wrong in the same way as with the minimal library; its coefficients are attenuated together, so it evolves several times too slowly, shows no intermittency, and settles at the wrong energy. Boundedness here is an observed property of these fitted models, not a guarantee for arbitrary coefficients in the overcomplete library. The zero-mean projection suppresses an otherwise undamped mean-vorticity component that spurious terms could drive.

\end{appendices}

% \newpage
\begingroup
\setlength{\bibsep}{0pt}
\setlength{\bibhang}{1em}
\fontsize{7pt}{7.5pt}\selectfont
\bibliographystyle{unsrtnat}
\bibliography{ref}
\endgroup
\end{document}